\documentclass[prd,preprint,showpacs]{revtex4}
\usepackage{chay,epsfig}
\begin{document}
\title{Gauge-invariant Lagrangian of the soft-collinear
   effective theory and its application to soft-collinear currents}
\author{Junegone Chay}\email{chay@korea.ac.kr}
\author{Chul Kim}\email{chulkim@korea.ac.kr}
\affiliation{Department of Physics, Korea University, Seoul 136-701,
Korea} 
\preprint{KUPT-04-01}
\begin{abstract}
We construct the soft-collinear effective Lagrangian 
which is manifestly gauge invariant order by order. Field
redefinitions of collinear gauge fields and a proper decomposition
of quark fields are necessary to make the Lagrangian gauge
invariant. We can obtain the effective Lagrangian order by order in 
$\mathrm{SCET}_{\mathrm{I}}$ and $\mathrm{SCET}_{\mathrm{II}}$ by
adopting appropriate power counting methods. Various types of
current operators can be investigated using this formalism, and we
choose the soft-collinear current as a specific example to 
present explicit radiative corrections. The
hard-collinear, collinear, and ultrasoft modes in
$\mathrm{SCET}_{\mathrm{I}}$ and the collinear and soft modes in
$\mathrm{SCET}_{\mathrm{II}}$ reproduce the infrared divergences of
the full theory and the matching can be performed. We show our results
explicitly using two types of regularization schemes. The first scheme is
the dimensional regularization both for the ultraviolet and the
infrared divergences with on-shell particles, and the second scheme is
regulating the infrared divergence with the off-shellness of external
particles. We discuss other types of operators in terms of the
two-step matching.
\end{abstract}


\maketitle

\section{Introduction}
Soft-collinear effective theory (SCET) is a useful tool to extract
important physics from a system with energetic light
quarks \cite{Bauer:2000yr, Bauer:2001ct, Bauer:2001yt, Chay:2002vy,
Bauer:2002nz,Bauer:2002uv,Beneke:2002ph,Beneke:2002ni}. In describing  
energetic light particles, we decompose the momentum of a collinear
particle as 
\begin{equation}
p^{\mu} =\frac{\overline{n}\cdot p}{2} n^{\mu} +p_{\perp}^{\mu} +
\frac{n\cdot p}{2} \overline{n}^{\mu},
\label{cmo}
\end{equation}
where $n^{\mu}$ is the light-like direction in which the collinear
particle moves, and $\overline{n}^{\mu}$ is a light-like vector
in the opposite direction, satisfying $n^2 = \overline{n}^2=0$, 
$n\cdot \overline{n}=2$. And $p_{\perp}^{\mu}$ is the momentum
perpendicular to $n^{\mu}$ and $\overline{n}^{\mu}$. Here the
component $\overline{n}\cdot p$ is the largest scale of order $Q$,
$p_{\perp}^{\mu}$ is the intermediate scale, and $n\cdot p$ is the
smallest component. 

For a system with energetic light quarks, the relevant scales are a
large energy $Q$ and the typical hadronic scale $\Lambda \sim
\Lambda_{\mathrm{QCD}}$, and we describe the power counting in terms
of the ratio of these two scales. We will eventually describe
the light quarks with $p^2 \sim \Lambda^2$, which is
appropriate for the quarks to form light hadrons. In this case, the
collinear momentum scales as 
$(\overline{n}\cdot p, p_{\perp}, n\cdot p)\sim (Q,\Lambda,
\Lambda^2/Q)$. However, we consider the intermediate effective theory,
called $\mathrm{SCET}_{\mathrm{I}}$, in which the momentum of the
collinear particles scales as $(\overline{n}\cdot p, p_{\perp}, n\cdot
p) \sim (Q,\sqrt{Q\Lambda}, \Lambda)$ such that $p^2 \sim
Q\Lambda$. In the literature \cite{Becher:2003qh,Beneke:2003pa}, these
modes are sometimes  labelled as ``hard-collinear'' modes. There are
soft particles with momentum $p_s^{\mu} \sim (\sqrt{Q\Lambda},
\sqrt{Q\Lambda}, \sqrt{Q\Lambda})$ in $\mathrm{SCET}_{\mathrm{I}}$,
but these particles do not interact with collinear particles, and we
do not consider them here. There are also ultrasoft (usoft) particles
with momentum $p_{us} \sim (\Lambda, \Lambda,\Lambda)$ with $p_{us}^2
\sim \Lambda^2$. The intermediate effective theory
$\mathrm{SCET}_{\mathrm{I}}$ can be obtained by integrating out the
hard degrees of freedom of order $Q$ from the full theory. Then we
obtain the final effective theory $\mathrm{SCET}_{\mathrm{II}}$ by
integrating out the degrees of freedom of order $p^2 \sim
Q\Lambda$. In $\mathrm{SCET}_{\mathrm{II}}$, the collinear particles
scale as $p_c \sim (Q,\Lambda,\Lambda^2/Q)$, and the soft momentum,
which was labelled as the usoft momentum in
$\mathrm{SCET}_{\mathrm{I}}$, now scales as $p_s \sim
(\Lambda,\Lambda,\Lambda)$. The small expansion parameter in SCET is
chosen as the ratio $p_{\perp}/\overline{n} \cdot p \sim
p_{\perp}/Q$. In $\mathrm{SCET}_{\mathrm{I}}$, the expansion parameter
is $\sqrt{\Lambda/Q}$, and in $\mathrm{SCET}_{\mathrm{II}}$, it is of
order $\Lambda/Q$.

This systematic method to obtain the effective theories is a two-step
matching \cite{Bauer:2003mg}. It is convenient and transparent to use
this two-step matching where there are some contributions of order
$p^2 =Q\Lambda$, as in the spectator contributions for $B$ decays
\cite{Bauer:2002aj,Pirjol:2002km,Chay:2003zp,Chay:2003ju,Chay:2003kb,
  Mantry:2003uz}. The 
construction of the operators, and the computation of the
hard-collinear contributions in $\mathrm{SCET}_{\mathrm{I}}$ can be
systematically performed, and we can
easily go down to $\mathrm{SCET}_{\mathrm{II}}$. On the other hand, we
can directly match the full theory to the final effective theory 
$\mathrm{SCET}_{\mathrm{II}}$, but proper care should be taken. This
is discussed in detail in Sec.~\ref{sec4}.

In constructing the effective Lagrangian in SCET, we require that the
gauge invariance under collinear and (u)soft gauge transformations
be preserved order by order. This has been partially accomplished in
Ref.~\cite{Bauer:2003mg} by treating the $n\cdot A_{us}$ component of
the usoft gauge fields as a background field for collinear gauge
transformations, and by redefining the collinear gauge field to
make the collinear Lagrangian gauge invariant order by order.
As a result, the collinear Lagrangian in $\mathrm{SCET}_{\mathrm{I}}$
is gauge invariant at each order, and the usoft-collinear Lagrangian
is also gauge invariant as a byproduct, but 
it is not clear whether the usoft Lagrangian at higher orders is gauge
invariant. 

In order to make the Lagrangian gauge invariant at each order, we
treat all the components of the usoft gauge field as background fields
for collinear gauge transformations, and we also redefine the
collinear gauge field accordingly. The fact that all the components of 
the usoft gauge field should act as background fields is also
important in factorizing the usoft contributions.  And when we
decompose the quark field in the full theory into the collinear quark
fields and the usoft quark field, we put an additional phase factor to
satisfy the transformation properties under the collinear and usoft
gauge transformations. Then the collinear gauge fields are redefined
by factorizing the usoft interactions in $\mathrm{SCET}_{\mathrm{I}}$ 
to obtain the gauge-invariant Lagrangian 
order by order in $\mathrm{SCET}_{\mathrm{I}}$ and we go down to
$\mathrm{SCET}_{\mathrm{II}}$. We 
employ the appropriate power counting methods in each effective theory
and list the first few terms of the Lagrangian in each sector. 

The main ingredient in constructing effective theories is that we modify
the ultraviolet behavior of the theory, but the infrared behavior
should be intact. This is physically
obvious since both theories share the same infrared region. It means
that if there is infrared divergence in the full theory, it should be
reproduced in the effective theory. Only after we assure that the
infrared behavior is the same in both theories, we can match the two
theories at the boundary, and compute the Wilson coefficients and the
anomalous dimensions of various
operators because the infrared divergences cancel in matching. 

Recently it has been argued that the collinear and the soft degrees of
freedom in the final effective theory $\mathrm{SCET}_{\mathrm{II}}$
may not be enough to reproduce the infrared divergence in the full
theory at one loop \cite{Becher:2003qh,Becher:2003kh}. In this paper, we
consider radiative corrections  
to soft-collinear current operators and show that the infrared
divergences in the full theory are fully reproduced both in
$\mathrm{SCET}_{\mathrm{I}}$ and $\mathrm{SCET}_{\mathrm{II}}$ using
the conventional hard-collinear, collinear, and usoft modes in
$\mathrm{SCET}_{\mathrm{I}}$  and the collinear, soft modes in
$\mathrm{SCET}_{\mathrm{II}}$. 

In classifying the collinear and the soft degrees of freedom in
$\mathrm{SCET}_{\mathrm{II}}$, note that the collinear momentum scales 
as $p_c \sim (Q,\Lambda, \Lambda^2/Q)$. For the momentum to be
collinear, the component $\overline{n}\cdot p_c$ must be of order $Q$,
not smaller than that, and the remaining components are determined
by requiring that $p_c^2\sim \Lambda^2$. However the soft momentum is
different. We generically say that the soft momentum scales as 
$p_s \sim (\Lambda,\Lambda,\Lambda)$, but it means that $\Lambda$ is
the maximum fluctuation. When we refer to soft dynamics, we consider
the dynamics with scales less than $\Lambda$. Therefore 
further classification of the soft momentum, in which
momenta with $p^2<\Lambda^2$ are labelled differently from the soft
momentum, may be conceptually helpful, but it is not necessary in the
construction of the effective theory and radiative corrections. 

In this paper, we consider a specific example of soft-collinear
current operators and compute their radiative corrections in the full
QCD, $\mathrm{SCET}_{\mathrm{I}}$ and
$\mathrm{SCET}_{\mathrm{II}}$. We can employ dimensional
regularization both for the ultraviolet and the infrared divergences
with external particles on their mass shell. In this case it is
difficult to extract the infrared divergent part alone. In order to
separate the infrared divergence from the ultraviolet divergence, we
use the off-shell regularization scheme for the infrared divergence
and the dimensional regularization for the ultraviolet
divergence. In this case, care must be taken in using the factorized
form for the usoft interactions. After we separate the infrared and
the ultraviolet divergences, we use the on-shell 
regularization scheme with the dimensional regularization both for the
ultraviolet and the infrared divergences in order to calculate the
anomalous dimensions of the soft-collinear current operators in each
effective theory.  

The organization of the paper is as follows: In Section \ref{sec2}, we
construct the effective Lagrangian in $\mathrm{SCET}_{\mathrm{I}}$,
which is gauge invariant under both collinear and usoft gauge
transformations at each order in $\sqrt{\Lambda/Q}$. 
And we evolve down to $\mathrm{SCET}_{\mathrm{II}}$ and construct the
effective Lagrangian in $\mathrm{SCET}_{\mathrm{II}}$ employing an
appropriate power counting. In 
Section \ref{sec3}, we construct the effective 
soft-collinear current operators in SCET and compute the radiative
corrections at one loop. We show that the infrared divergences in the
full QCD are the same both in $\mathrm{SCET}_{\mathrm{I}}$ and in
$\mathrm{SCET}_{\mathrm{II}}$, and compute the Wilson coefficients and
the anomalous dimension of the soft-collinear current operator. In
Sec.~\ref{sec4}, we compare the scaling behavior of the soft-collinear
current with that of the back-to-back collinear current and the 
heavy-to-light current. Finally in Sec.~\ref{sec5} we present a
conclusion. 

\section{\label{sec2} Effective Lagrangian in SCET}
\subsection{Effective Lagrangian in $\mathrm{SCET}_{\mathrm{I}}$}
The effective Lagrangian in $\mathrm{SCET}_{\mathrm{I}}$ is derived
by integrating out the hard degrees of freedom of order $Q$ in the
full theory. In $\mathrm{SCET}_{\mathrm{I}}$ there are the collinear
modes with $p^2 \sim Q\Lambda$ and the usoft modes with $p^2\sim
\Lambda^2$. We decompose the gauge field $A^{\mu}$ in the full QCD
into the collinear gauge field $A_n^{\mu}$, the soft field
$A_s^{\mu}$, and the usoft field $A_{us}^{\mu}$. The collinear gauge
field scales as $A_n^{\mu}= (\overline{n}\cdot A_n, A_{n\perp}^{\mu},
n\cdot A_n)\sim (Q,\sqrt{Q\Lambda}, \Lambda)$, 
the soft gauge field scales as $A_s^{\mu} \sim (\sqrt{Q\Lambda},
\sqrt{Q\Lambda}, \sqrt{Q\Lambda})$ and the usoft field scales as
$A_{us}^{\mu} \sim (\Lambda, \Lambda, \Lambda)$. Since the soft fields
do not couple to the collinear sector, we will not consider them from
now on.  

Let us consider the gauge symmetries in $\mathrm{SCET}_{\mathrm{I}}$
starting from the full QCD. The full QCD possesses an $SU(3)$ color
gauge symmetry, and the gauge field $A^{\mu}$ transforms as
\begin{equation}
A^{\mu} \rightarrow UA^{\mu} U^{\dagger} +\frac{i}{g}
U[\partial^{\mu} U^{\dagger}], 
\end{equation}
where $U$ is an $SU(3)$ gauge transformation and the bracket means
that the differential operator $\partial^{\mu}$ acts only inside the
bracket. In SCET, there are 
collinear, soft and usoft gauge transformations, which are subsets of
the gauge transformations of the full theory. The collinear (usoft) gauge
transformation $U_c$ ($U_{us}$) is the gauge transformation which
satisfies $\partial^{\mu} U_c \sim (Q,\sqrt{Q\Lambda},\Lambda)$ 
[$\partial^{\mu} U_{us} \sim (\Lambda, \Lambda,\Lambda)$]. We
decompose the gauge field $A^{\mu}$ as
\begin{equation}
A^{\mu} = A_n^{\mu} + A_{us}^{\mu},
\end{equation}
and we extract the label momentum from the collinear gauge field
$A_n^{\mu}$  as
\begin{equation}
A_n^{\mu} (x) =\sum_q e^{-iq\cdot x} A_{n,q}^{\mu} (x),
\end{equation}
where $q^{\mu} = \overline{n}\cdot q n^{\mu}/2 +q_{\perp}^{\mu}
=\mathcal{O} (Q) + \mathcal{O} (\sqrt{Q\Lambda})$ is the label
momentum, and the derivative acting on $A_{n,q}^{\mu}(x)$ gives the
momentum of order $\Lambda$. We also decompose the collinear gauge
transformation as 
\begin{equation}
U_c =\sum_p e^{-ip\cdot x} \mathcal{U}_{c,p}.
\end{equation}
Since the large label momentum is extracted, the derivative
acting on $\mathcal{U}_{c,p}$ gives $\partial^{\mu} \mathcal{U}_{c,p}
\sim \Lambda$. From now on, we will drop the label index and the
summation over the label momentum unless they are necessary. 

Since the fluctuation of the usoft gauge field is of order
$1/\Lambda$, the variation is smooth compared to that of 
the collinear fields. Therefore the usoft gauge field $A_{us}^{\mu}$
acts as a classical background field under collinear gauge
transformations, and $A_n^{\mu}$ transforms homogeneously under usoft
gauge transformations. That is, the gauge fields transform as
\cite{Abbott:1980hw} 
\begin{equation}
A_n^{\mu} \rightarrow \mathcal{U}_c (A_n^{\mu} +A_{us}^{\mu})
\mathcal{U}_c^{\dagger} 
+\frac{1}{g} \mathcal{U}_c [(\mathcal{P}^{\mu} +i\partial^{\mu})
  \mathcal{U}_c^{\dagger}] -A_{us}^{\mu}, \ \ 
A_{us}^{\mu} \rightarrow A_{us}^{\mu},
\end{equation}
under collinear gauge transformations, where $\mathcal{P}^{\mu}$ is
the label momentum operator with $n\cdot \mathcal{P}=0$.
Under usoft gauge transformations, the gauge fields transform as
\begin{equation}
A_n^{\mu} \rightarrow U_{us} A_n^{\mu} U_{us}^{\dagger}, \ \
A_{us}^{\mu} \rightarrow U_{us} A_{us} U_{us}^{\dagger} +\frac{i}{g}
U_{us} [\partial^{\mu} U_{us}^{\dagger}].
\end{equation}

As a first step toward the gauge invariant formulation of the
Lagrangian order by order, we redefine the collinear gauge fields as  
\begin{equation}
gA_n^{\mu}= g\hat{A}_n^{\mu} +\hat{W} [iD_{us}^{\mu},
  \hat{W}^{\dagger}],
\label{newa}
\end{equation}
where $iD_{us}^{\mu} = i\partial_{us}^{\mu} + gA_{us}^{\mu}$. The
derivative $\partial^{\mu}$ is written as $\partial^{\mu}
=\partial_c^{\mu} +\partial_{us}^{\mu}$, where $\partial_c^{\mu}$
($\partial_{us}^{\mu}$) picks out collinear (usoft) momentum of order
$\Lambda$ from collinear (usoft) fields, and gives zero when applied to
usoft (collinear) fields. In
$\mathrm{SCET}_{\mathrm{I}}$, $i\partial_c^{\mu}$ is 
negligible compared to $\mathcal{P}^{\mu}$, but it is kept to make
the transition to $\mathrm{SCET}_{\mathrm{II}}$ transparent.

The hatted fields $\hat{A}_n^{\mu}$ are newly defined fields and
$\hat{W}$ is given by \cite{Beneke:2002ph,Bauer:2003mg}
\begin{equation}
\hat{W} =WZ^{\dagger} = \Bigl[ \sum_{\mathrm{perm}} \exp
  \Bigl(-\frac{g}{\overline{\mathcal{P}}} \overline{n} \cdot \hat{A}_n
  \Bigr) \Bigr],
\end{equation}
where $\overline{\mathcal{P}} = \overline{n} \cdot \mathcal{P}$, and
the Wilson lines $W$ and $Z$ are defined as 
\begin{eqnarray}
W&=& P \exp \Bigl( ig \int_{-\infty}^0 ds \overline{n} \cdot (A_n
+A_{us}) (\overline{n} s+x) \Bigr), \nonumber \\
Z&=& P \exp \Bigl( ig \int_{-\infty}^0 ds \overline{n} \cdot 
A_{us} (\overline{n} s+x) \Bigr).
\end{eqnarray}
The hatted gauge field $\hat{A}_n^{\mu}$ transforms under 
collinear gauge transformations as
\begin{equation}
g\hat{A}_n^{\mu} \rightarrow \mathcal{U}_c
        g\hat{A}_n^{\mu}\mathcal{U}_c^{\dagger} 
        +\mathcal{U}_c  [(\mathcal{P}^{\mu} +i\partial^{\mu}_c)
          \mathcal{U}_c^{\dagger} ],
\label{chaykim}
\end{equation}
which shows that $\hat{A}_n^{\mu}$ transforms like a collinear gauge
field under collinear gauge transformations, and the transformation
does not involve any usoft gauge fields. 

The definition of $\hat{A}_n^{\mu}$ is similar to the definition in
Ref.~\cite{Bauer:2003mg}, but in their definition, $n\cdot
\hat{A}_n(=n\cdot A_n) $ does not change, and $\hat{A}_n^{\mu}$
transforms as 
\begin{equation}
g\hat{A}_n^{\mu} \rightarrow \mathcal{U}_c g\hat{A}_n^{\mu}
\mathcal{U}_c^{\dagger} +\mathcal{U}_c \Bigl[\overline{n}\cdot
  \mathcal{P} \frac{n^{\mu}}{2} +\mathcal{P}_{\perp}^{\mu} +in\cdot
  D_{us} \frac{\overline{n}^{\mu}}{2}, \mathcal{U}_c^{\dagger}\Bigr],
\label{bauer}
\end{equation}
which still involves usoft gauge fields. In some sense, this
definition looks reasonable because only the $n\cdot A_n$ component is
of order $\Lambda$, while other components such as $\overline{n}\cdot
A_n$ and $A_{n\perp}^{\mu}$ are larger than $\Lambda$. Therefore we
may decompose the fields of order $\Lambda$ into the quantum field
$n\cdot A_n$ and the background field $n\cdot A_{us}$ for
collinear gauge transformations. However, small fluctuations of order
$\Lambda$ are allowed for the components $\overline{n}\cdot A_n$ and
$A_{n\perp}^{\mu}$ though the characteristic scales are of order $Q$,
and $\sqrt{Q\Lambda}$ respectively. Therefore we can put
$\overline{n}\cdot A_{us}$ and $A_{us,\perp}^{\mu}$ as background
fields for $\overline{n}\cdot A_n$ and $A_{n\perp}^{\mu}$ for those
small fluctuations. From our definition, the usoft gauge
field truly becomes a background field for collinear gauge
transformations, and it does not spoil the power counting. 
Furthermore, it is critical to keep all the components of 
the usoft gauge field as background fields for collinear gauge
transformations in showing the decoupling
of the usoft interactions with the collinear particles and in writing
the gauge fixing term and the ghost Lagrangian, in which the collinear
and the usoft modes are decoupled.

\begin{table}[b]
\begin{tabular}{c|cc}\hline
Fields&Collinear $\mathcal{U}_c$&Usoft $U_{us}$ \\ \hline
$\xi_n$& $\mathcal{U}_c \xi_n$& $U_{us} \xi_n$\\ 
$\xi_{\overline{n}}$& $\mathcal{U}_c \xi_{\overline{n}} $& $U_{us}
\xi_{\overline{n}}$ \\ 
$gA_n^{\mu}$ & $\mathcal{U}_c g(A_n^{\mu} + A_{us}^{\mu})
\mathcal{U}_c^{\dagger} -gA_{us}^{\mu} +\mathcal{U}_c
        [(\mathcal{P}^{\mu} +i\partial^{\mu}) \mathcal{U}_c^{\dagger}
        ]$& $U_{us}A_n^{\mu} U_{us}^{\dagger}$ \\
$g\hat{A}_n^{\mu}$ & $\mathcal{U}_c
        g\hat{A}_n^{\mu}\mathcal{U}_c^{\dagger} 
        +\mathcal{U}_c  [(\mathcal{P}^{\mu} +i\partial^{\mu}_c)
          \mathcal{U}_c^{\dagger} ]$& $U_{us}\hat{A}_n^{\mu}
        U_{us}^{\dagger}$ \\ 
$q_{us}$& $q_{us}$& $U_{us}q_{us}$ \\
$gA_{us}^{\mu}$& $gA_{us}$& $U_{us} gA_{us} U_{us}^{\dagger} + U_{us}
        [i\partial_{us} U_{us}^{\dagger}]$ \\
$\hat{W}$& $\mathcal{U}_c \hat{W}$& $U_{us} \hat{W} U_{us}^{\dagger}$
\\ 
$Y$& $Y$& $U_{us}Y$ \\ \hline
\end{tabular}
\caption{\label{table1}Gauge transformation properties of the
  quantities in $\mathrm{SCET}_{\mathrm{I}}$ under collinear and usoft
  gauge transformations.}
\end{table}

In deriving the effective Lagrangian for the quark sector, we also
decompose the quark field $\psi$ in the full QCD into the usoft field
$q_{us}$ and the collinear fields $\xi_{n,p}$ and
$\xi_{\overline{n},p}$. We express the collinear fields extracting the
label momentum as $\sum_p e^{-ip\cdot x} q_{n,p} (x)$,
and $\xi_{n,p} =(\FMslash{n} \FMslash{\overline{n}}/4) q_{n,p}$,
$\xi_{\overline{n},p} =(\FMslash{\overline{n}}\FMslash{n}/4)
q_{n,p}$. We also drop the label index $p$ and the summation from now on.
The collinear quark fields $\xi_n$, $\xi_{\overline{n}}$ transform
under collinear and usoft gauge transformations as
\begin{eqnarray}
\mathrm{collinear}:&&\xi_n \rightarrow \mathcal{U}_c \xi_n, \ \
\xi_{\overline{n}} 
\rightarrow \mathcal{U}_c \xi_{\overline{n}}, \ \ q_{us} \rightarrow
q_{us} \nonumber \\
\mathrm{usoft}:&&\xi_n \rightarrow U_{us} \xi_n, \ \ \xi_{\overline{n}}
\rightarrow U_{us} \xi_{\overline{n}}, \ \ q_{us} \rightarrow U_{us}
q_{us}. 
\end{eqnarray}
The gauge transformation properties of the gauge fields and the quark
fields in $\mathrm{SCET}_{\mathrm{I}}$ are summarized in
Table~\ref{table1}. 

At first sight, we may decompose the quark field as 
\begin{equation}
\psi = \xi_n + \xi_{\overline{n}} + q_{us},
\label{decom}
\end{equation}
but this decomposition is too naive. In general, the usoft field can
have an additional phase factor. In order to determine the relative
phase of $q_{us}$ with respect to the collinear fields, note that the
quark field $\psi$ and the covariant derivative $iD^{\mu}
=i\partial^{\mu} +gA^{\mu} $ in the full QCD  transform as
\begin{equation}
\psi \rightarrow U\psi, \ iD^{\mu} \rightarrow UiD^{\mu} U^{\dagger},
\ iD^{\mu} \psi \rightarrow UiD^{\mu} \psi.
\label{fcov}
\end{equation}
Since the collinear and usoft gauge transformations are subsets of the
full gauge transformations, $(iD_c^{\mu} +iD_{us}^{\mu})\psi$  should
transform as in Eq.~(\ref{fcov}) with $U$ replaced by $U_c$ or
$U_{us}$. (Here we can neglect the soft modes since it does not change
the result.) We find that the combination
\begin{equation}
\psi = q_c + \hat{W} q_{us} = \xi_n + \xi_{\overline{n}} + \hat{W}
q_{us}
\label{trued} 
\end{equation}
satisfies this property. With the transformation properties in
Table~\ref{table1},  the sum of the covariant
derivatives $iD_c^{\mu} + iD_{us}^{\mu}$ transforms homogeneously under
collinear and usoft gauge transformations as
\begin{equation}
iD_c^{\mu} + iD_{us}^{\mu} \rightarrow \left\{ \begin{array}{cc}
\mathcal{U}_c (iD_c^{\mu} + iD_{us}^{\mu}) \mathcal{U}_c^{\dagger}, &
\mathrm{collinear}, \\
U_{us} (iD_c^{\mu} + iD_{us}^{\mu}) U_{us}^{\dagger}, & \mathrm{usoft}.
                                               \end{array}
\right.
\end{equation}
Then it follows that
\begin{equation}
(iD_c^{\mu} + iD_{us}^{\mu}) (q_c +\hat{W} q_{us}) \rightarrow \left\{
  \begin{array}{cc} 
\mathcal{U}_c (iD_c^{\mu} + iD_{us}^{\mu}) (q_c +\hat{W} q_{us}), &
\mathrm{collinear},\\ 
U_{us} (iD_c^{\mu} + iD_{us}^{\mu}) (q_c +\hat{W} q_{us}), &
\mathrm{usoft},
  \end{array}
\right.
\end{equation}
as desired.

Now the starting point to derive the effective Lagrangian from the
full QCD Lagrangian $\mathcal{L} =\overline{\psi} i\FMSlash{D} \psi$
reads as
\begin{eqnarray}
\mathcal{L} &=& \overline{\xi}_n \frac{\FMslash{\overline{n}}}{2}
in\cdot D\xi_n +\overline{\xi}_n i\FMSlash{D}_{\perp}
\xi_{\overline{n}} +\overline{\xi}_n i\FMSlash{D} \hat{W} q_{us}
+\overline{q}_{us} \hat{W}^{\dagger} i\FMSlash{D} \xi_n
+\overline{q}_{us} \hat{W}^{\dagger} i\FMSlash{D} \xi_{\overline{n}}
\nonumber \\
&+& \overline{q}_{us} \hat{W}^{\dagger} i\FMSlash{D} \hat{W} q_{us}
+\Bigl[ \overline{\xi}_{\overline{n}} i\FMSlash{D}_{\perp} \xi_n
  +\overline{\xi}_{\overline{n}} \frac{\FMslash{n}}{2} i\overline{n}
  \cdot D \xi_{\overline{n}} +\overline{\xi}_{\overline{n}}
  i\FMSlash{D} \hat{W} q_{us} \Bigr],
\label{startl}
\end{eqnarray}
and the equation of motion for $\xi_{\overline{n}}$ now becomes
\begin{equation}
\xi_{\overline{n}} = -\frac{1}{i\overline{n} \cdot D}
\frac{\FMslash{\overline{n}}}{2} (i\FMSlash{D}_{\perp} \xi_n +
i\FMSlash{D} \hat{W} q_{us}),
\end{equation}
which satisfies the collinear gauge transformation property in
Table~\ref{table1}. If there is no $\hat{W}$ in front of $q_{us}$,
$\xi_{\overline{n}}$ does not transform as $\xi_{\overline{n}}
\rightarrow \mathcal{U}_c \xi_{\overline{n}}$. 

Before we proceed further, it would be useful to compare our approach
with the formulations by Bauer et al. \cite{Bauer:2003mg} and by
Beneke et al. \cite{Beneke:2002ph,Beneke:2002ni}. In
Ref.~\cite{Bauer:2003mg}, they use the collinear gauge transformation,
Eq.~(\ref{bauer}), in which only the $n\cdot A_{us}$ acts as a
classical background field. They redefine $\hat{A}_n^{\mu}$ as shown
in Eq.~(\ref{newa}) except that $n\cdot \hat{A}_n = n\cdot A_n$. It
makes the collinear Lagrangian gauge invariant order by order. But
they do not introduce the relative phase $\hat{W}$ in front of
$q_{us}$ when the quark field in the full theory is decomposed in
$\mathrm{SCET}_{\mathrm{I}}$, which does not make the usoft Lagrangian
gauge invariant at higher orders. Beneke et
al. \cite{Beneke:2002ph,Beneke:2002ni} start from the same collinear
gauge transformation properties as ours, in 
which all the the usoft field is a background field. And they also
introduce the relative phase $\hat{W}$ in front of $q_{us}$. Then the
collinear quark and gluon fields are redefined to satisfy the
homogenized gauge transformation given by Eq.~(\ref{bauer}). The
redefined collinear fields involve Wilson lines made of usoft
gluons after choosing a light-cone gauge. The $\overline{n}\cdot A_n$
component of the collinear gauge field remains unchanged, while the
$n\cdot A_n$ component does not change in Ref.~\cite{Bauer:2003mg}. 

Here we treat the usoft field as a background field for collinear
gauge transformations, and put an additional phase $\hat{W}$ as in
Ref.~\cite{Beneke:2002ph,Beneke:2002ni}. Then we redefine the
collinear gauge field as in Eq.~(\ref{newa}) as a first step to make
the Lagrangian invariant order by order. And we express the Lagrangian
in terms of the hatted collinear gauge field, $\hat{W}$, and the
unchanged collinear and usoft quark fields.

The Lagrangian can be decomposed into the collinear, the
usoft-collinear, and the usoft Lagrangian $\mathcal{L}
=\mathcal{L}_c +\mathcal{L}_{\xi q}+\mathcal{L}_{us}$, depending on
the quark fields, and can be written as
\begin{eqnarray}
\mathcal{L}_c &=& \overline{\xi} \frac{\FMslash{\overline{n}}}{2}
\Bigl( in\cdot D + i\FMSlash{D}_{\perp} \frac{1}{i\overline{n} \cdot
  D} i\FMSlash{D}_{\perp} \Bigr) \xi, \nonumber \\
\mathcal{L}_{\xi q} &=& \overline{\xi} i\FMSlash{D} \hat{W} q_{us}
+\overline{\xi} \frac{\FMslash{\overline{n}}}{2} i\FMSlash{D}_{\perp}
\frac{1}{i\overline{n} \cdot D} i\FMSlash{D} \hat{W} q_{us}
+\mathrm{h.c.}, 
\nonumber \\
\mathcal{L}_{us} &=&\overline{q}_{us} \hat{W}^{\dagger} i\FMSlash{D}
\hat{W} q_{us} 
-\overline{q}_{us} \hat{W}^{\dagger} i\FMSlash{D}
\frac{1}{i\overline{n} \cdot D} \frac{\FMslash{\overline{n}}}{2}
i\FMSlash{D} \hat{W} q_{us},
\end{eqnarray}
where we also drop the index $n$ for the collinear quark field to
simplify the notation. In terms of the hatted field, the covariant
derivative is written as 
\begin{equation}
iD^{\mu} = iD_c^{\mu} + iD_{us}^{\mu} = i\hat{D}_c^{\mu} + \hat{W}
iD_{us}^{\mu} \hat{W}^{\dagger},
\end{equation}
where $i\hat{D}_c^{\mu} = \mathcal{P}^{\mu} +i\partial_c^{\mu} +
g\hat{A}_n^{\mu}$, and $iD_{us}^{\mu} = i\partial_{us}^{\mu}
+gA_{us}^{\mu}$.  

Before we go down to $\mathrm{SCET}_{\mathrm{II}}$, we decouple the
usoft interactions with the collinear particles by factorizing the
usoft interactions. This is achieved by the field redefinitions
\cite{Bauer:2001yt} 
\begin{equation}
\xi = Y\xi^{(0)}, \  \hat{A}_n^{\mu} =
Y\hat{A}^{\mu, (0)}_n Y^{\dagger},
\label{newy}
\end{equation}
where the Wilson line $Y$ is given by
\begin{equation}
Y(x) = P \exp \Bigl( ig\int_{-\infty}^0 ds n\cdot A_{us} (ns+x)
\Bigr).
\end{equation}
This is the second and final redefinition of the collinear fields, and
these fields will be matched to the corresponding fields in
$\mathrm{SCET}_{\mathrm{II}}$. In $\mathrm{SCET}_{\mathrm{I}}$, the
usoft interactions with collinear particles do not put the
intermediate states off the mass shell, hence this factorization does
not correspond to integrating out off-shell modes. However in
$\mathrm{SCET}_{\mathrm{II}}$, the corresponding soft interactions put
the intermediate collinear particles off the mass shell by $p^2\sim
Q\Lambda$, therefore these should be integrated out in
$\mathrm{SCET}_{\mathrm{II}}$. This is easily achieved by replacing
$Y$ by $S$, where the Wilson line $S$ is obtained by replacing
$A_{us}^{\mu}$ of $Y$ in $\mathrm{SCET}_{\mathrm{I}}$ by $A_s^{\mu}$
in $\mathrm{SCET}_{\mathrm{II}}$. 

In Ref.~\cite{Bauer:2001yt}, the usoft factorization for the collinear
fields is illustrated nicely by attaching usoft gluons to collinear
quarks and gluons. Here the attached usoft gluons are regarded as
background fields and when we sum over the radiation of usoft gluons
to all orders, it gives a path-ordered exponential $Y$, which depends
only on $n\cdot A_{us}$. This eikonal summation is possible only when
all the components of the usoft gluons are treated as background
fields. Therefore our treatment of the usoft field as a background
field for collinear gauge transformations is also justified in
factorizing the usoft interactions.

The fields $\xi^{(0)}$ and $\hat{A}_n^{\mu,(0)}$ are the fundamental
objects and they transform in a peculiar way under gauge
transformations. Under collinear gauge transformations, $\xi^{(0)}$
and $g\hat{A}_n^{\mu,(0)}$ transform as
\begin{eqnarray}
\xi^{(0)} &\rightarrow& Y^{\dagger} \mathcal{U}_c \xi = (Y^{\dagger}
\mathcal{U}_c Y )\xi^{(0)}, \nonumber \\
g\hat{A}_n^{\mu,(0)} &\rightarrow& Y^{\dagger} \Bigl(\mathcal{U}_c
  g\hat{A}_n^{\mu} \mathcal{U}_c^{\dagger} +\mathcal{U}_c
  [(\mathcal{P}^{\mu} +i\partial_c^{\mu})\mathcal{U}_c^{\dagger}
  ]\Bigr) Y \nonumber \\
&=& (Y^{\dagger} \mathcal{U}_c Y) g\hat{A}_n^{\mu,(0)}
   (Y^{\dagger} \mathcal{U}_c Y)^{\dagger}+
 (Y^{\dagger} \mathcal{U}_c Y)   \Bigl[(\mathcal{P}^{\mu}
    +i\partial_c^{\mu}) (Y^{\dagger} \mathcal{U}_c Y)^{\dagger} \Bigr],
\end{eqnarray}
where the last equality is obtained, using the fact that $Y$ commutes
with the collinear momentum operator $\mathcal{P}^{\mu}$ and
$i\partial_c^{\mu}$. Under usoft gauge transformations, they transform
as  
\begin{eqnarray}
\xi^{(0)} &\rightarrow& Y^{\dagger} U_{us}^{\dagger} U_{us} \xi =
\xi^{(0)}, \nonumber \\
g\hat{A}_n^{\mu,(0)} &\rightarrow& g\hat{A}_n^{\mu,(0)}.
\end{eqnarray}
Therefore $\xi^{(0)}$ and $\hat{A}_n^{\mu,(0)}$ transform under the
modified collinear gauge transformations $Y^{\dagger} \mathcal{U}_c Y$
as $\xi$ and $\hat{A}_n^{\mu}$ transform under the collinear 
gauge transformations $\mathcal{U}_c$, while they do not transform at
all under usoft gauge transformations. On the other hand, the usoft
fields $q_{us}$ and $A_{us}^{\mu}$ do not transform under the modified
collinear transformations. Therefore the collinear fields transform
only under collinear transformation, while the usoft fields transform
only under usoft gauge transformations. This makes the construction of
the gauge fixing terms simple. 

The Lagrangian for the gauge fields is given by
\begin{eqnarray}
\mathcal{L}_g &=& \frac{1}{2g^2} \mathrm{Tr} \Bigl( [iD_{\mu},
  iD_{\nu}] [iD^{\mu}, iD^{\nu}]\Bigr) \nonumber \\
&=& \frac{1}{2g^2} \mathrm{Tr} \Bigl( [i\hat{D}_c^{\mu} +
  \hat{W}Y^{\dagger}  iD_{us}^{\mu} Y \hat{W}^{\dagger},
  i\hat{D}_c^{\nu}  +\hat{W} Y^{\dagger} iD_{us}^{\nu}
  Y\hat{W}^{\dagger}]^2   \Bigr)\nonumber \\ 
&=& \frac{1}{2g^2} \mathrm{Tr} \Bigl( [i\hat{D}_c^{\mu},
i\hat{D}_c^{\nu}]\Bigr)^2 + \frac{1}{2g^2} \mathrm{Tr} \Bigl(
  [iD_{us}^{\mu}, iD_{us}^{\nu}]\Bigr)^2 +\cdots,
\end{eqnarray}
where the dots represent the interaction between collinear and usoft
gluons and we can systematically expand the Lagrangian in powers of
$\sqrt{\Lambda}$. Since the collinear fields and the usoft fields
transform separately under collinear and usoft gauge transformations,
it is simple to fix the gauge and to construct the ghost
Lagrangian. We can choose the gauge fixing term as
\begin{equation}
\mathcal{L}_{\mathrm{fix}} = -\frac{1}{2\alpha_i} (F_i^a)^2, \
(i=c,us), 
\end{equation}
where
\begin{equation}
F_c^a = -i(\mathcal{P}^{\mu} +i\partial_c^{\mu} )
\hat{A}_{n\mu}^{(0),a}, \ \  
F_{us}^a = \partial_{us}^{\mu} A_{us,\mu}^a.
\end{equation}
And the ghost Lagrangian becomes
\begin{eqnarray}
\mathcal{L}_{\mathrm{ghost}} &=& \overline{\eta}^a
\Bigl((\mathcal{P}^{\mu}+i\partial^{\mu}_c)^2  \delta^{ac} +igf^{abc}
(\mathcal{P}^{\mu}+i\partial^{\mu}_c)  \hat{A}_{n\mu}^{(0),a} \Bigr)
\eta^c \nonumber \\
&+& \overline{c}^a (-\partial_{us}^2 \delta^{ac} -gf^{abc}
\partial_{us}^{\mu} A_{us,\mu}^b) c^c,
\end{eqnarray}
where $\eta^a$ ($c^a$) is the collinear (usoft) ghost field. 

According to the transformation Eq.~(\ref{newy}), we have $\hat{W} = Y
\hat{W}^{(0)} Y^{\dagger}$, and from now on, we will drop the
superscript ``(0)''. The effective Lagrangian can be expanded 
Lagrangian in powers of $\sqrt{\Lambda/Q}$. The first three terms of
the collinear Lagrangian are given by 
\begin{eqnarray}
\mathcal{L}_c^{(0)} &=& \overline{\xi}
\frac{\FMslash{\overline{n}}}{2} \Bigl( in\cdot \hat{D}_c + 
i\hat{\FMSlash{D}}_{c\perp} \frac{1}{i\overline{n} \cdot \hat{D}_c}
i\hat{\FMSlash{D}}_{c\perp} \Bigr) \xi, \nonumber \\
\mathcal{L}_c^{(1)} &=& \overline{\xi}
\frac{\FMslash{\overline{n}}}{2} \Bigl( i\hat{\FMSlash{D}}_{c\perp}
\frac{1}{i\overline{n} \cdot \hat{D}_c} \hat{W}
Y^{\dagger} i\FMSlash{D}_{us}^{\perp} Y\hat{W}^{\dagger} + \hat{W}
Y^{\dagger} i\FMSlash{D}_{us}^{\perp} Y\hat{W}^{\dagger}
\frac{1}{i\overline{n} 
  \cdot \hat{D}_c}  i\hat{\FMSlash{D}}_{c\perp} \Bigr) \xi, \nonumber
\\
\mathcal{L}_c^{(2)} &=& \overline{\xi}
\frac{\FMslash{\overline{n}}}{2} \Bigl( \hat{W}
Y^{\dagger} i\FMSlash{D}_{us}^{\perp} Y \hat{W}^{\dagger}
\frac{1}{i\overline{n}   \cdot \hat{D}_c}   \hat{W} Y^{\dagger}
i\FMSlash{D}_{us}^{\perp} Y \hat{W}^{\dagger} \nonumber \\
&&- i\hat{\FMSlash{D}}_{c\perp} \frac{1}{i\overline{n} \cdot
  \hat{D}_c} \hat{W} Y^{\dagger} i\overline{n}\cdot D_{us} Y
\hat{W}^{\dagger} \frac{1}{i\overline{n} \cdot   \hat{D}_c}
i\hat{\FMSlash{D}}_{c\perp} \Bigr) \xi.
\end{eqnarray}
In deriving $\mathcal{L}_c^{(0)}$, we use the following relations
\begin{eqnarray}
\overline{\xi} \frac{\FMslash{\overline{n}}}{2}   in\cdot
  D  \xi &=&\overline{\xi} \frac{\FMslash{\overline{n}}}{2}  \Bigl( in\cdot
  \hat{D}_c +\hat{W} in\cdot D_{us}
  \hat{W}^{\dagger} \Bigr) \xi \nonumber \\
&\rightarrow& \overline{\xi}  \frac{\FMslash{\overline{n}}}{2}
  Y^{\dagger} \Bigl( i n\cdot \partial_c 
   +Y gn\cdot \hat{A}_n Y^{\dagger} 
  +Y\hat{W} Y^{\dagger} in\cdot D_{us} Y\hat{W}^{\dagger} Y^{\dagger}
  \Bigr) Y\xi \nonumber \\
&=& \overline{\xi}  \frac{\FMslash{\overline{n}}}{2} \Bigl( in\cdot
  \hat{D}_c +\hat{W} Y^{\dagger} in\cdot D_{us} Y \hat{W} \Bigr) \xi
= \overline{\xi}  \frac{\FMslash{\overline{n}}}{2} i n\cdot \hat{D}_c
  \xi, 
\end{eqnarray}
where we use the fact that $in\cdot D_{us} Y = Yin\cdot
\partial_{us}$, and the usoft derivative operator applied to the
collinear fields in $\hat{W}$ and $\xi$ yields zero.

The usoft-collinear Lagrangian is given by
\begin{eqnarray}
\mathcal{L}_{\xi q}^{(1)} &=& \overline{\xi}
i\hat{\FMSlash{D}}_c^{\perp} \hat{W} Y^{\dagger} q_{us} +
\mathrm{h.c.}, \nonumber \\
\mathcal{L}_{\xi q}^{(2)} &=& \overline{\xi}
\frac{\FMslash{\overline{n}}}{2} \Bigl( in\cdot \hat{D}_c
+i\hat{\FMSlash{D}}_c^{\perp} \frac{1}{i\overline{n} \cdot \hat{D}_c}
i\hat{\FMSlash{D}}_c^{\perp} \Bigr) \hat{W} Y^{\dagger} q_{us}
+\mathrm{h.c.}, 
\end{eqnarray}
and the usoft quark Lagrangian is written as
\begin{eqnarray}
\mathcal{L}_{us}^{(0)} &=& \overline{q}_{us} i\FMSlash{D}_{us} q_{us},
\nonumber \\
\mathcal{L}_{us}^{(3)} &=& \overline{q}_{us} Y \hat{W}^{\dagger}
i\hat{\FMSlash{D}}_c^{\perp} \hat{W} Y^{\dagger} q_{us}, \nonumber \\
\mathcal{L}_{us}^{(4)} &=& \overline{q}_{us}
\frac{\FMslash{\overline{n}}}{2}  Y \hat{W}^{\dagger}  \Bigl( in\cdot
\hat{D}_c +i\hat{\FMSlash{D}}_c^{\perp} \frac{1}{i\overline{n} \cdot
  \hat{D}_c} i\hat{\FMSlash{D}}_c^{\perp} \Bigr) \hat{W} Y^{\dagger}
q_{us}. 
\end{eqnarray}
In the present form, the effective Lagrangian at each order is
manifestly gauge invariant under collinear and usoft gauge
transformations. The importance of the relative phase $\hat{W}$ of
$q_{us}$ in decomposing the quark fields is clearly seen in
$\mathcal{L}_{us}^{(3), (4)}$. If there were no relative phase
$\hat{W}$, these Lagrangians would not be gauge invariant, and the
gauge invariance can be sustained only after we include $\hat{W}$ in
front of $q_{us}$.

\subsection{Effective Lagrangian in $\mathrm{SCET}_{\mathrm{II}}$}

In $\mathrm{SCET}_{\mathrm{II}}$, we rename the usoft fields as the
soft fields and we will replace $Y$ by $S$. We can write down the
effective Lagrangian in $\mathrm{SCET}_{\mathrm{II}}$ in powers of
$\Lambda$. However, there is change in organizing the Lagrangian
because the power counting of the fields changes. In
$\mathrm{SCET}_{\mathrm{II}}$, the collinear field $\xi$ scales as
$\sim \Lambda$, the soft quark field $q_s$ (also the heavy quark field $h$
if any) scales as $\sim
\Lambda^{3/2}$. For the collinear gauge field, it scales as
$\hat{A}_n^{\mu} \sim (Q, \Lambda, \Lambda^2/Q)$, and for the soft gauge
field $A_s^{\mu} \sim (\Lambda,\Lambda,\Lambda)$. In the action, the
power counting for the integration measure $d^4 x$ also changes. For
the Lagrangian which involves collinear fields only or soft fields
only, the integration measure becomes $d^4 x \sim \Lambda^{-4}$, but
if collinear fields and soft fields are mixed, the integration measure
scales as $d^4 x\sim \Lambda^{-3}$
\cite{Hill:2002vw,Stewart:2003gt}. With this in mind, the effective
Lagrangian in 
$\mathrm{SCET}_{\mathrm{II}}$, which is gauge invariant under
collinear and soft gauge transformations, can be organized in powers
of $\Lambda$. 

The collinear quark Lagrangian is given as
\begin{eqnarray}
\mathcal{L}_{c,\mathrm{II}}^{(0)} &=& \overline{\xi} \Bigl(
in\cdot \hat{D}_c +i\hat{\FMSlash{D}}_{c\perp}
\hat{W} \frac{1}{\overline{\mathcal{P}}} \hat{W}^{\dagger}
i\hat{\FMSlash{D}}_{c\perp} \Bigr) \frac{\FMslash{\overline{n}}}{2}
\xi, \nonumber \\
\mathcal{L}_{c,\mathrm{II}}^{(1)} &=& (\overline{\xi} \hat{W})
(S^{\dagger} i\FMSlash{D}_s^{\perp} S)
\frac{1}{\overline{\mathcal{P}}} (\hat{W}^{\dagger}
i\hat{\FMSlash{D}}_{c\perp} \frac{\FMslash{\overline{n}}}{2} \xi) +
(\overline{\xi} i\hat{\FMSlash{D}}_{c\perp} \hat{W})
\frac{1}{\overline{\mathcal{P}}} (S^{\dagger}
i\FMSlash{D}_s^{\perp} S)  \frac{\FMslash{\overline{n}}}{2}
(\hat{W}^{\dagger} \xi) \nonumber \\
&&+ (\overline{\xi} \hat{W}) S^{\dagger} i\FMSlash{D}_s^{\perp}
\frac{1}{\overline{\mathcal{P}}} i\FMSlash{D}_s^{\perp} S
\frac{\FMslash{\overline{n}}}{2} (\hat{W}^{\dagger} \xi).
\label{c2}
\end{eqnarray}
Note that the third term in $\mathcal{L}_{c,\mathrm{II}}^{(1)}$
previously belonged to $\mathcal{L}_c^{(2)}$ in
$\mathrm{SCET}_{\mathrm{I}}$, but 
now it is included in $\mathcal{L}_{c,\mathrm{II}}^{(1)}$ due to the
different power counting. The
soft-collinear quark Lagrangian is given by
\begin{eqnarray}
\mathcal{L}_{sc,\mathrm{II}}^{(1/2)} &=& (\overline{\xi} \hat{W} )
(\hat{W}^{\dagger} i\hat{\FMSlash{D}}_{c\perp} \hat{W}) S^{\dagger}
q_s +\mathrm{h.c.}, \nonumber \\
\mathcal{L}_{sc, \mathrm{II}}^{(3/2)} &=& (\overline{\xi} \hat{W}) 
\frac{\FMslash{\overline{n}}}{2} \Bigl( \hat{W}^{\dagger} in\cdot
\hat{D}_c \hat{W} 
+ \hat{W}^{\dagger} i\hat{\FMSlash{D}}_{c\perp} \hat{W}
\frac{1}{\overline{\mathcal{P}}} 
\hat{W}^{\dagger}i\hat{\FMSlash{D}}_{c\perp} \hat{W} \nonumber \\
&+&\hat{W}^{\dagger} i\FMSlash{D}_c^{\perp} \hat{W}^{\dagger}
\frac{1}{\overline{\mathcal{P}}}  S^{\dagger}
i\FMSlash{D}_s^{\perp} S + S^{\dagger}i\FMSlash{D}_s^{\perp} S
\frac{1}{\overline{\mathcal{P}}} \hat{W}^{\dagger}
i\FMSlash{D}_c^{\perp} \hat{W}^{\dagger}  \Bigr) S^{\dagger} q_s.
\label{sc2}
\end{eqnarray}
Note that the superscripts have changed reflecting that
$\mathcal{L}_{sc,\mathrm{II}}^{(1/2)}$ is suppressed by
$\Lambda^{1/2}$ compared to $\mathcal{L}_c^{(0)}$, and the orders are
rearranged with the power counting $iD_c^{\perp}\sim iD_s^{\perp}
\sim \Lambda$. The soft Lagrangian is given by
\begin{equation}
\mathcal{L}_{s,\mathrm{II}}^{(0)} = \overline{q}_s i\FMSlash{D}_s q_s, \
\mathcal{L}_{s,\mathrm{II}}^{(1)} = \overline{q}_s S\hat{W}^{\dagger}
i\FMSlash{D}_c^{\perp} \hat{W} S^{\dagger} q_s.
\label{s2}
\end{equation}
The higher-order terms in the Lagrangian in
$\mathrm{SCET}_{\mathrm{I}}$ are redistributed according to the power
counting rules in $\mathrm{SCET}_{\mathrm{II}}$. And there can be
many-quark operators by integrating out the off-shell fields, which we
do not present here. Formally in order to obtain the effective
Lagrangian in $\mathrm{SCET}_{\mathrm{II}}$, we introduce
hard-collinear auxiliary
fields, which scale as $(Q,\sqrt{Q\Lambda},\Lambda)$, and they are
integrated out. The effective action in $\mathrm{SCET}_{\mathrm{II}}$
is given by 
\begin{equation}
e^{i\Gamma_{\mathrm{II},\mathrm{eff}}} = \int \mathcal{D} \xi_X
  \mathcal{D} \overline{\xi}_X \mathcal{D} A_{nX} e^{iS_{\mathrm{I}}
  (\xi+\xi_X +\overline{\xi}+\overline{\xi}_X,
  A_n^{\mu}+A_{nX}^{\mu},\mathrm{soft fields})} ,
\end{equation}
where $S_{\mathrm{I}}$ is the action in $\mathrm{SCET}_{\mathrm{I}}$
and all the fields with the subscript $X$ are the auxiliary fields of
order $\sim (Q,\sqrt{Q\Lambda},\Lambda)$, which are to be integrated
out.

The effective Lagrangians in Eqs.~(\ref{c2})-- (\ref{s2}) 
are the effective Lagrangian in
$\mathrm{SCET}_{\mathrm{II}}$. However, care must be taken in deriving
Feynman rules if we expand each Lagrangian in powers of
$g$. Especially when the Wilson lines $\hat{W}$ and $S$ are expanded,
we have to take the momentum conservation into account.  
Note that in the soft-collinear interactions, there should be more
than two collinear particles and two soft particles to conserve
momentum. For example, consider
$\mathcal{L}_{sc,\mathrm{II}}^{(1/2)}$. At leading order in $g$, it 
contains a collinear quark $\xi$, a collinear gluon $A_n^{\mu}$ and a
soft quark $q_{us}$. But this is not allowed since the momentum is not
conserved. Though we add any number of collinear momenta in the
$n^{\mu}$ direction, we cannot make a soft momentum. Therefore the
nonzero leading term in $\mathcal{L}_{sc,\mathrm{II}}^{(1/2)}$ starts
from order $g^2$, in which a soft gluon is added by expanding $S$,
conserving momentum. In $\mathcal{L}_{s,\mathrm{II}}^{(1)}$, it
starts with a single collinear gluon, which is not allowed either, and  
there should be at least two collinear gluons.

The selecting procedure described above is somewhat cumbersome, and it
would be desirable to express the Lagrangian in which the gauge 
invariance and the momentum conservation are shown manifestly. One way
to accomplish this is to choose a specific gauge, in which the
Lagrangian takes a simpler form. For example, if 
we choose the gauge $\hat{W}=S=1$ \cite{Beneke:2002ni}, all the
multi-fields from 
the Wilson lines vanish and we can easily impose the momentum
conservation. In $\mathcal{L}_{c,\mathrm{II}}^{(1)}$,
only the last term survives. No terms survive in
$\mathcal{L}_{sc,\mathrm{II}}^{(1/2)}$, and only the last two terms
survive in $\mathcal{L}_{sc,\mathrm{II}}^{(3/2)}$. In the soft
Lagrangian, $\mathcal{L}_{s,\mathrm{II}}^{(1)}$ vanishes. The
construction of the gauge-invariant and momentum-conserving Lagrangian 
in $\mathrm{SCET}_{\mathrm{II}}$ without choosing a specific gauge,
and the comparison with the Lagrangian with a specific gauge will be
discussed elsewhere. 
 
\section{\label{sec3}Soft-collinear current}
As a specific example, we consider the radiative corrections for the
soft-collinear current of the form $\overline{\xi} \Gamma q_{(u)s}$ from
the current $\overline{\psi}\Gamma \psi$ in the full theory, where
$\Gamma$ is a Dirac matrix. The
usoft-collinear current operators in $\mathrm{SCET}_{\mathrm{I}}$ can be
obtained by replacing the quark field in the full theory by
Eq.~(\ref{trued}), and it is given by
\begin{equation}
\overline{q} \Gamma q \rightarrow (\overline{\xi}_n +
\overline{\xi}_{\overline{n}} +\overline{q}_{us} \hat{W}^{\dagger} )
\Gamma (\xi_n +\xi_{\overline{n}} +\hat{W} q_{us}).
\end{equation}
 Using the equation of motion for $\xi_{\overline{n}}$, and expanding
 the currents in powers of $\sqrt{\Lambda}$, we can obtain the current
 operators in $\mathrm{SCET}_{\mathrm{I}}$. We list the first few
 current operators at tree level after we factorize
 the usoft interactions by redefining $\xi \rightarrow Y\xi$,
 $\hat{A}_n^{\mu} \rightarrow Y \hat{A}_n^{\mu} Y^{\dagger}$. The
 collinear-collinear current operators are given by
\begin{eqnarray}
J_c^{(0)} &=& \overline{\xi} \Gamma \xi, \nonumber \\
J_c^{(1)} &=& \overline{\xi} \Bigl(\frac{\FMslash{\overline{n}}}{2}
i\FMSlash{\hat{D}}_c^{\perp} \frac{1}{i\overline{n} \cdot \hat{D}_c}
\Gamma  +\Gamma \frac{1}{i\overline{n} \cdot \hat{D}_c}
i\FMSlash{\hat{D}}_c^{\perp} \frac{\FMslash{\overline{n}}}{2} \Bigr)
\xi, \nonumber \\
J_c^{(2)} &=&\overline{\xi} \Bigl( i\FMSlash{\hat{D}}_c^{\perp}
\frac{\FMslash{\overline{n}}}{2} \frac{1}{i\overline{n} \cdot
  \hat{D}_c} \Gamma \frac{1}{i\overline{n} \cdot \hat{D}_c}
\frac{\FMslash{\overline{n}}}{2}  i\FMSlash{\hat{D}}_c^{\perp}
\nonumber \\
&+& \frac{\FMslash{\overline{n}}}{2} \hat{W} Y^{\dagger}
i\FMSlash{D}_{us}^{\perp} Y \hat{W}^{\dagger} \frac{1}{i\overline{n}
  \cdot \hat{D}_c} \Gamma +  \Gamma \frac{1}{i\overline{n} 
  \cdot \hat{D}_c} \hat{W} Y^{\dagger}
i\FMSlash{D}_{us}^{\perp} Y \hat{W}^{\dagger}
\frac{\FMslash{\overline{n}}}{2} \Bigr)\xi,
\label{jc}
\end{eqnarray}
and the usoft-collinear current operators are given by
\begin{eqnarray}
J_{\xi q}^{(2)} &=& \overline{\xi}  \hat{W} \Gamma Y^{\dagger} q_{us}
+\mbox{h.c.}, 
\nonumber \\
J_{\xi q}^{(3)} &=& -\overline{\xi} \Bigl( i\FMSlash{D}_c^{\perp}
\frac{\FMslash{\overline{n}}}{2}  \frac{1}{i\overline{n} \cdot
  \hat{D}_c} \Gamma + \Gamma \frac{1}{i\overline{n} \cdot  \hat{D}_c}
\frac{\FMslash{\overline{n}}}{2} i\FMSlash{D}_c^{\perp} \Bigr) \hat{W}
Y^{\dagger} q_{us} + \mbox{h.c.},
\label{jsc}
\end{eqnarray} 
and the usoft current operators are given by
\begin{eqnarray}
J_{us}^{(4)} &=& \overline{q}_{us} \Gamma q_{us}, \nonumber \\
J_{us}^{(5)} &=& -\overline{q}_{us} \hat{W}^{\dagger} \Bigl( \Gamma
\frac{1}{i\overline{n} \cdot   \hat{D}_c} \frac{\FMslash{\overline{n}}}{2}
i\FMSlash{D}_c^{\perp} + i\FMSlash{D}_c^{\perp}
\frac{\FMslash{\overline{n}}}{2} \frac{1}{i\overline{n} \cdot
  \hat{D}_c} \Gamma \Bigr) \hat{W} q_{us}.
\label{jus}
\end{eqnarray}
Note that the power counting of these operators is performed in
$\mathrm{SCET}_{\mathrm{I}}$, that is, in powers of
$\sqrt{\Lambda/Q}$, and the superscripts in the current operators show
the power dependence with respect to $J_c^{(0)}$. When we go down to
$\mathrm{SCET}_{\mathrm{II}}$, $Y$ should be replaced by $S$, and the
power counting in $\mathrm{SCET}_{\mathrm{II}}$ should be applied. In
$\mathrm{SCET}_{\mathrm{I}}$, 
$iD_c^{\mu} \sim (Q,\sqrt{Q\Lambda},\Lambda)$, $iD_{us} \sim (\Lambda,
\Lambda, \Lambda)$, but in $\mathrm{SCET}_{\mathrm{II}}$, $iD_c^{\mu}
\sim (Q,\Lambda,\Lambda^2/Q)$, $iD_{us} \sim (\Lambda, 
\Lambda, \Lambda)$. Therefore $iD_c^{\perp} \sim iD_{us}^{\perp}$ in
$\mathrm{SCET}_{\mathrm{II}}$, while they have different power
behavior in $\mathrm{SCET}_{\mathrm{I}}$. As an example, the first
term in $J_c^{(2)}$ is suppressed compared to the other terms in
$\mathrm{SCET}_{\mathrm{II}}$.

We compute the vertex corrections at one loop. We consider a special
case with $\Gamma=\gamma^{\mu}$ at the end of the computation, but we
keep $\Gamma$ when the results hold for an arbitrary $\Gamma$. 
In our computation, two regularization
schemes are employed. The first one is the on-shell regularization 
in which the external particles are on their mass shell, and we use
the dimensional regularization both for the infrared and the
ultraviolet divergences. However, there can be cancellation between
the infrared and the ultraviolet divergences in this scheme. This
happens especially in $\mathrm{SCET}_{\mathrm{II}}$ and all the
radiative corrections are zero due to this cancellation. In order to
separate the infrared and the ultraviolet divergences, we use
the off-shell regularization
scheme in which the infrared divergence is regulated by the
off-shellness of the external particles and we use the dimensional
regularization for the ultraviolet divergence. The computation in this
scheme is more complicated, but we can explicitly check if the
infrared divergences are reproduced in the effective theories. After
we show that the infrared divergences of the full theory are
reproduced in $\mathrm{SCET}_{\mathrm{I}}$ and
$\mathrm{SCET}_{\mathrm{II}}$, we can match the full theory and SCET
since the infrared divergence is canceled in matching. We match the
full theory and $\mathrm{SCET}_{\mathrm{I}}$ at $\mu \sim Q$, and
match $\mathrm{SCET}_{\mathrm{I}}$ and $\mathrm{SCET}_{\mathrm{II}}$
at $\mu\sim \sqrt{Q\Lambda}$, using the on-shell regularization
scheme to compute the Wilson coefficients and the anomalous dimensions
of the (u)soft-collinear current operator.  

\subsection{Full QCD calculation}

\begin{figure}[t]
\begin{center}
\epsfig{file=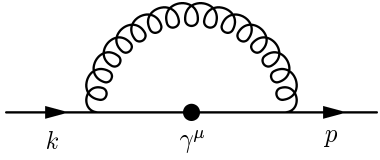, width=4.5cm}
\end{center}
\vspace{-1.5cm}
\caption{Vertex correction to the soft-collinear current in the full
  theory.}
\label{fig1}
\end{figure}

In full QCD, we consider the vertex correction to the soft-collinear
current, which is shown in Fig.~\ref{fig1}. The Feynman graph yields
\begin{equation}
M_F = -ig^2 C_F \int \frac{d^D l}{(2\pi)^D} \frac{\gamma_{\alpha}
  (\FMslash{l} +\FMslash{p}) \gamma_{\mu} (\FMslash{l} +\FMslash{k})
  \gamma^{\alpha}}{l^2 (l+p)^2 (l+k)^2},
\label{mf}
\end{equation}
where $D=4-2\epsilon$, $C_F = (N^2-1)/(2N)$ is the $SU(3)$ color
factor with $N=3$. 

If we put the external particles on the mass shell ($k^2=p^2=0$), and
use the dimensional regularization both for the infrared and the
ultraviolet divergences, the one-loop correction in this scheme is
given by 
\begin{equation}
M_F^D = \frac{\alpha_s C_F}{4\pi} \gamma_{\mu} I_F (q^2),
\label{mfd}
\end{equation}
where  $q^2 = (p-k)^2=-2 p\cdot k$, and $I_F (q^2)$ is given as 
\begin{eqnarray}
I_F (q^2)&=&\Bigl( \frac{-q^2}{\mu^2}
\Bigr)^{-\epsilon} \Bigl(\frac{1}{\epsilon_{\mathrm{UV}}}
-\frac{2}{\epsilon^2} 
-\frac{4}{\epsilon} -8+\frac{\pi^2}{6} \Bigr) \nonumber \\
&=& \frac{1}{\epsilon_{\mathrm{UV}}} 
-\frac{2}{\epsilon^2} +\Bigl( -4 +2\ln \frac{-q^2}{\mu^2}
\Bigr)\frac{1}{\epsilon} +3 \ln \frac{-q^2}{\mu^2} 
-\ln^2
\frac{-q^2}{\mu^2}-8+\frac{\pi^2}{6}.
\end{eqnarray}
The pole $1/\epsilon_{\mathrm{UV}}$ comes from
the ultraviolet divergence, which will be canceled by the wave function
renormalization due to the current conservation, and other
$1/\epsilon$ poles are of the infrared origin. 

If we put the external particles off the mass shell $p^2,\ k^2 \neq
0$, the calculation is more complicated, but the graph in
Fig.~\ref{fig1} gives 
\begin{equation}
M_F^C =\frac{\alpha_s C_F}{4\pi} \gamma_{\mu} \Bigl(
\frac{1}{\epsilon_{\mathrm{UV}}} -  \ln \frac{-q^2}{\mu^2}-2 \ln
\frac{p^2}{q^2} \ln \frac{k^2}{q^2} -2 \ln \frac{p^2}{q^2} -2 \ln
\frac{k^2}{q^2} -\frac{2\pi^2}{3} \Bigr),
\label{irf}
\end{equation}
where the pole $1/\epsilon_{\mathrm{UV}}$ is the ultraviolet
divergence, and the infrared divergences are regulated by the
off-shellness. Here $q^2 = (p-k)^2 = -2p\cdot k +p^2 +k^2$, but $p^2$
and $k^2$ can be neglected compared to $p\cdot k$.    

\subsection{Calculation in SCET}
In $\mathrm{SCET}_{\mathrm{I}}$, the calculation of the vertex
correction is more complicated, but it can be performed
systematically. In constructing the effective Lagrangian in
$\mathrm{SCET}_{\mathrm{I}}$, we have not distinguished the collinear
modes with $p^{\mu}_c \sim (Q,\Lambda, \Lambda^2/Q)$ and the
hard-collinear modes with $p^{\mu}_{hc} \sim (Q,\sqrt{Q\Lambda},
\Lambda)$. However, it is conceptually convenient to distinguish
these two contributions because the hard-collinear
contributions will be integrated out in order to obtain
$\mathrm{SCET}_{\mathrm{II}}$, but the collinear contributions still
remain in $\mathrm{SCET}_{\mathrm{II}}$. 
This point becomes transparent when 
we use the off-shellness of the external particles as an infrared
cutoff in the computation. If we have a loop momentum
$l^{\mu}$ and an external collinear momentum $p^{\mu}\sim
(Q,\Lambda,\Lambda^2/Q)$, there is a propagator whose denominator is
given by $(l+p)^2$. When the loop momentum is hard-collinear with
$l^{\mu} \sim (Q,\sqrt{Q\Lambda}, \Lambda)$, it is given by
\begin{equation}
(l+p)^2 = l^2 +\overline{n}\cdot p n\cdot l + (p^2 + n\cdot p
  \overline{n}  \cdot l + 2l_{\perp}\cdot p_{\perp}) = \mathcal{O}
  (Q\Lambda) + \mathcal{O} (Q\Lambda) +\mathcal{O}
  (Q\Lambda\sqrt{\frac{\Lambda}{Q}}). 
\label{hc}
\end{equation}
On the other hand, if the loop momentum is collinear with $l^{\mu} \sim
(Q,\Lambda, \Lambda^2/Q)$, it becomes
\begin{equation}
(l+p)^2 = l^2 +\overline{n}\cdot p n\cdot l +(p^2 + n\cdot p
  \overline{n}  \cdot l + 2l_{\perp}\cdot p_{\perp}) =
  \mathcal{O}  (\Lambda^2) + \mathcal{O} (\Lambda^2) +\mathcal{O}
  (\Lambda^2).
\label{co} 
\end{equation}
Therefore when the loop momentum $l^{\mu}$ is hard-collinear, we do
not have to introduce the infrared cutoff since the denominator can
never reach the infrared region near $\Lambda^2$. But when $l^{\mu}$
is collinear, the integral can be infrared divergent from this
propagator, and we need an infrared cutoff to regulate the infrared
divergence.

\begin{figure}[b]
\begin{center}
\vspace{0.5cm}
\epsfig{file=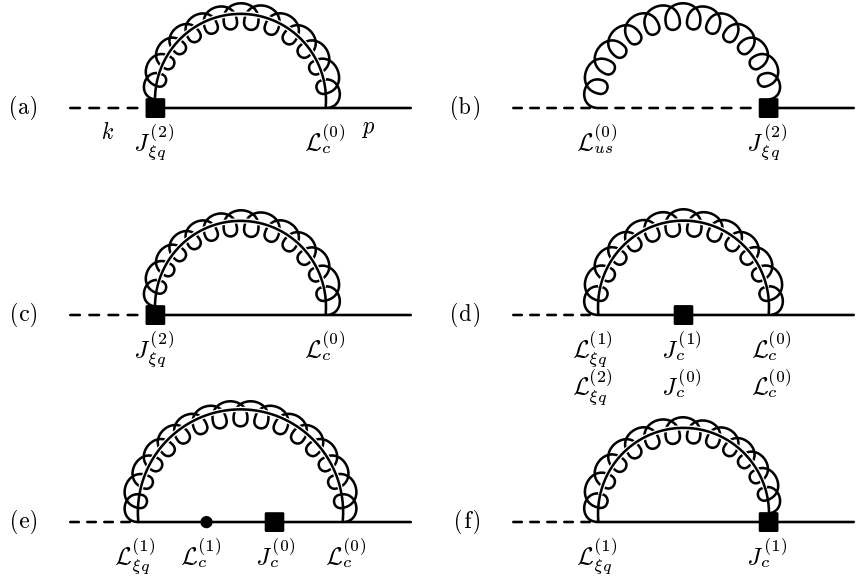, width=10.5cm}
\end{center}
\vspace{-1.0cm}
\caption{Vertex correction to the soft-collinear current in
$\mathrm{SCET}_{\mathrm{I}}$. The square box denotes the  inserted
current operators. The curly line with a solid line in (a) is a
collinear gluon, and the curly line in (b) is a soft gluon. The curly
line with a solid line in (c)--(f) is a hard-collinear gluon. 
The solid line is a collinear quark and the dashed line is an usoft
quark.}
\label{fig2}
\end{figure}

In $\mathrm{SCET}_{\mathrm{I}}$, we consider the contributions from
the collinear modes with $p^{\mu}_c \sim (Q,\Lambda,
\Lambda^2/Q)$, the usoft modes with $p^{\mu}_{us} \sim
(\Lambda,\Lambda, \Lambda)$, and the hard-collinear modes with
$p^{\mu}_{hc} \sim (Q,\sqrt{Q\Lambda}, \Lambda)$. 
The Feynman diagrams for the vertex correction are shown in
Fig.~\ref{fig2}. Fig.~\ref{fig2} (a) shows the collinear contribution, 
in which the loop momentum is collinear,
and Fig.~\ref{fig2} (b) is the usoft contribution. The remaining four
diagrams Fig.~\ref{fig2} (c) -- (f) are the hard-collinear
contributions. The insertions of the current operators and the
Lagrangian are explicitly shown in the figure. 
We express the results using the off-shellness $(p^2,\
k^2 \neq 0$) in order to see if the infrared divergence in the full
theory in Eq.~(\ref{irf}) is reproduced in
$\mathrm{SCET}_{\mathrm{I}}$. 

Before we compute the radiative corrections in Fig.~\ref{fig2}, let us
consider the propagators which appear in the calculation. The external
collinear particle has the collinear momentum 
$p^{\mu}$ which scales as $p^{\mu} \sim (Q,\Lambda,\Lambda^2/Q)$ with
$p^2 \sim \Lambda^2$. The usoft momentum $k^{\mu}$ scales as $k^{\mu}
\sim (\Lambda,\Lambda,\Lambda)$ with $k^2 \sim \Lambda^2$. The
propagator of the collinear quark is given from $\mathcal{L}_c^{(0)}$
as
\begin{equation}
i\frac{\FMslash{n}}{2} \frac{1}{n\cdot p
  +p_{\perp}^2/\overline{n}\cdot p} = i\frac{\FMslash{n}}{2}
  \frac{\overline{n} \cdot p}{p^2}.
\end{equation}
And the internal quark propagators in Fig.~\ref{fig2} have $(l+p)^2$
or $(l+k)^2$ in the denominator.  
In the collinear contribution as in Fig.~\ref{fig2} (a), $l^{\mu} \sim
(Q,\Lambda, \Lambda^2/Q)$, and all the terms in $(l+p)^2$  are of the
same order and it should be written as $(l+p)^2$ according to
Eq.~(\ref{co}). Here the off-shellness $p^2$ acts as an infrared
cutoff. In the usoft contribution of Fig.~\ref{fig2} (b), $l^{\mu}
\sim (\Lambda, \Lambda, \Lambda)$, and all the terms in $(l+k)^2$ are
of the same order and it should be written as $(l+k)^2$ and $k^2$ acts
as an infrared cutoff. In the hard-collinear contributions as in
Fig.~\ref{fig2} (c) -- (f), the loop momentum $l^{\mu}$ is
hard-collinear, scaling as $l^{\mu} \sim (Q,\sqrt{Q\Lambda}, \Lambda)$
with $l^2 \sim Q\Lambda$. In this case, $(l+p)^2$ can be written as
\begin{equation}
(l+p)^2 = l^2 + n\cdot l \overline{n} \cdot p +\mathcal{O} (Q\Lambda
\sqrt{\frac{\Lambda}{Q}}),  
\label{copo}
\end{equation}
where the first two terms are at leading order $\mathcal{O}
(Q\Lambda)$, and there is no need for the infrared cutoff $p^2$ since
the leading term can never reach the region $\sim \Lambda^2$.  
Similarly, $(l+k)^2$ can be written as 
\begin{equation}
(l+k)^2 \approx l^2 + \overline{n} \cdot l n\cdot k +\mathcal{O}
 (Q\Lambda \sqrt{\frac{\Lambda}{Q}}),  
\end{equation}
where $l^2 \approx \overline{n} \cdot l n\cdot k \sim Q\Lambda$. 
Therefore the infrared cutoff $k^2$ does not have to be kept in the
denominator either for the hard-collinear contributions.

Note that the off-shellness $p^2$ is of order $\Lambda^2$  for
external collinear particles. However, since the off-shellness 
$p^2$ acts only as an infrared cutoff, we can relax the off-shellness
as $p^2 <\Lambda^2$ as long as $\overline{n} \cdot p \sim Q$ is fixed.
As long as $\overline{n}\cdot p\sim
Q$, we can use the power counting as in Eq.~(\ref{copo}). Since $p^2$
is an infrared cutoff, we can make it as small as we want and we can
show that the infrared divergences are fully reproduced using any
small $p^2$, and there is no need to include any other degrees of
freedom corresponding to smaller $p^2$. 

\begin{figure}[t]
\begin{center}
\epsfig{file=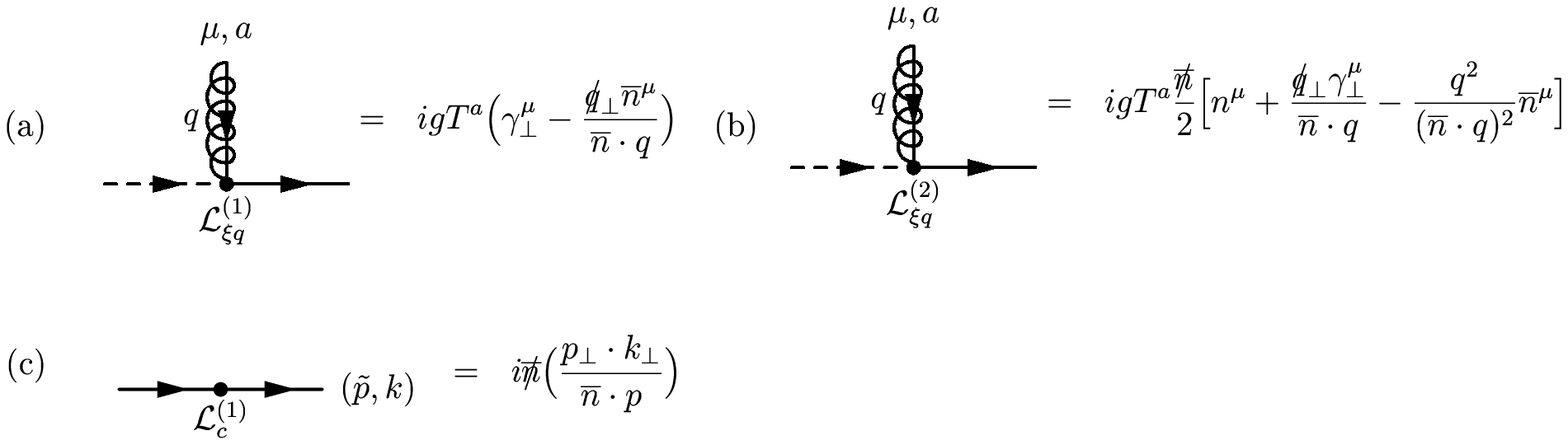, width=16.0cm}
\vspace{-0.8cm}
\end{center}
\caption{Feynman rules for the subleading Lagrangian
  necessary for the one-loop corrections to the soft collinear
  current. The momentum $q^{\mu}$ is incoming.} 
\label{fig3}
\end{figure}

\begin{figure}[b]
\begin{center}
\epsfig{file=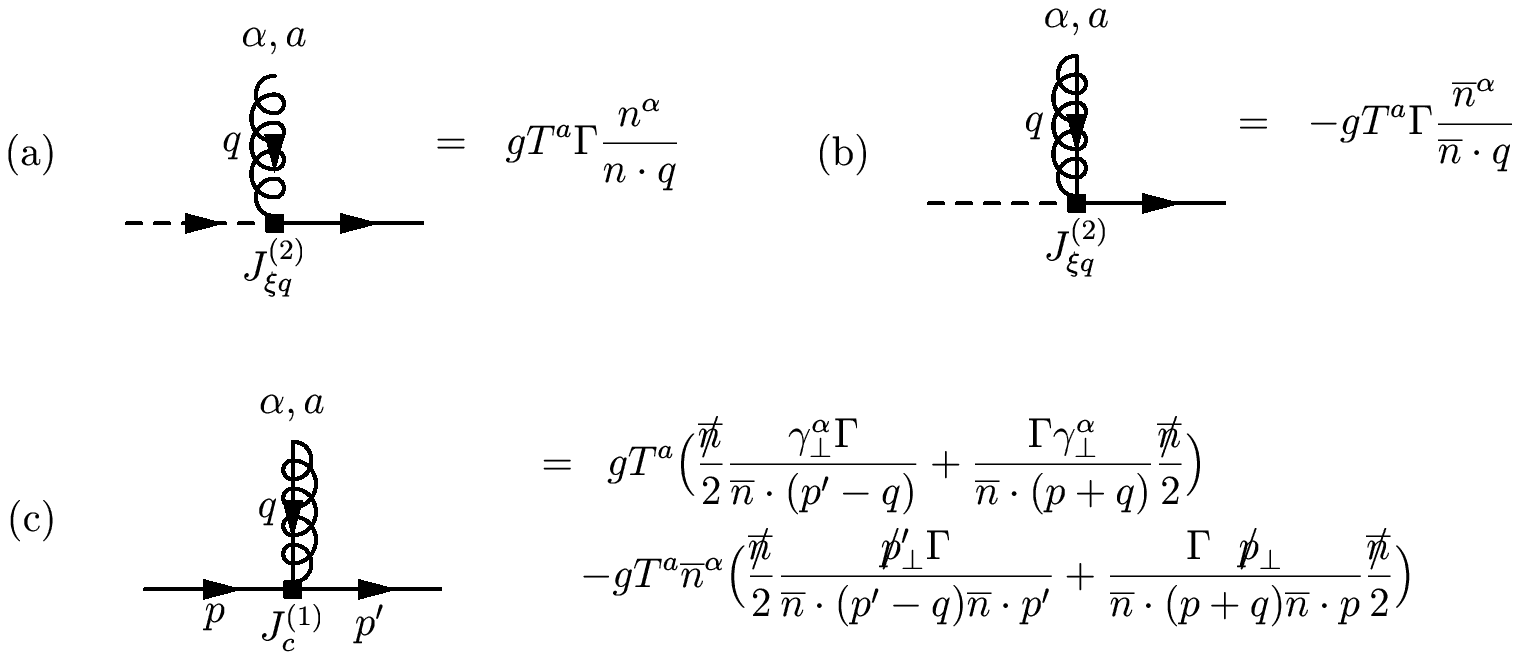, width=14.0cm}
\vspace{-0.8cm}
\end{center}
\caption{Feynman rules for the current operators necessary
for the one-loop corrections to the soft collinear current. The
momentum $q^{\mu}$ is incoming.}
\label{fig4}
\end{figure}

\begin{figure}[t]
\begin{center}
\epsfig{file=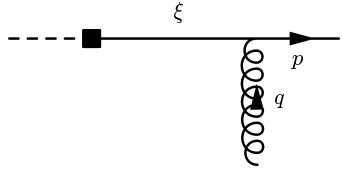, width=4.0cm}
\vspace{-0.7cm}
\end{center}
\caption{Single-gluon vertex for the usoft-collinear current operator
  in $\mathrm{SCET}_{\mathrm{I}}$ when the external particle is off
  the mass shell. The usoft momentum $q$ is incoming.} 
\label{fig5}
\end{figure}

The Feynman rules for the subleading Lagrangian, which appears in
Fig.~\ref{fig2},  are listed in
Fig.~\ref{fig3}. The Feynman rules for the currents can be derived from
Eqs.~(\ref{jc}), (\ref{jsc}) and (\ref{jus}), and we list those
necessary for the one-loop corrections of the soft-collinear current
in Fig.~\ref{fig4}. Note that the Feynman
rule in Fig.~\ref{fig4} (a) for the usoft-collinear current with an
usoft gluon is derived
from the current $\overline{\xi} \hat{W} \Gamma
Y^{\dagger} q_{us}$, in which the usoft Wilson line $Y$ is obtained by
putting the external particles on the mass shell. Therefore we have to
modify the Feynman rules for the usoft-collinear current when we put
the particles off the mass shell to extract the infrared divergence
separately.  The vertex for the usoft-collinear current with a single
usoft gluon in $\mathrm{SCET}_{\mathrm{I}}$ with the off-shell
external particles can be computed from the Feynman diagram in
Fig.~\ref{fig5}, and the amplitude is given as 
\begin{equation}
\overline{\xi} (ign\cdot A_{us} \frac{\FMslash{\overline{n}}}{2}) i 
\frac{\FMslash{n}}{2} \frac{\overline{n} \cdot (p-q)}{(p-q)^2} 
\Gamma q_{us} \rightarrow -g\overline{\xi} 
\frac{n\cdot A_{us}\overline{n} \cdot p}{-\overline{n} \cdot p n\cdot
  q +p^2}  \Gamma q_{us},
\label{modus}
\end{equation}
where $q^{\mu}$ is the incoming usoft momentum, and the leading
terms are kept after the arrow. 

In the denominator of Eq.~(\ref{modus}), the maximum fluctuation of
$\overline{n}\cdot p 
n\cdot q$ is of order $Q\Lambda$, but it can approach $\Lambda^2$ or
smaller since $q$ is usoft. Therefore we need an infrared cutoff $p^2$
and both terms should be kept. We can derive the Feynman rule for the
vertex of the usoft-collinear current with an usoft gluon from
Eq.~(\ref{modus}). When we put $p^2=0$, we obtain the Feynman rule in
Fig.~\ref{fig3} (a). At higher orders in 
$\alpha_s$, the modified Feynman rules may be complicated, but it can
be derived at each order. Bauer et al. \cite{Bauer:2003td} have
considered another form of the regulator in
$\mathrm{SCET}_{\mathrm{II}}$ to all orders in 
$\alpha_s$. However, the Feynman rules for the off-shell particles are
useful in distinguishing and separating the infrared divergence from a
given Feynman diagram. Once we know that the infrared divergence of
the full theory is reproduced in SCET, we go back to the dimensional
regularization with on-shell particles. So we do not dwell further on
the possible construction of the modified Feynman rules for the
off-shell particles. 

We do not have to modify the Feynman rule in Fig.~\ref{fig4} (b) for
the collinear gluon from expanding $\hat{W}$ when the external
particles are off the mass shell. In this case, we consider the
interaction of an usoft quark and a collinear quark emitting a
collinear gluon. In this case the momentum squared of the intermediate
collinear quark becomes $(q+k)^2 = \overline{n}\cdot q n\cdot k +
\mathcal{O} (\Lambda^2)$, where $q^{\mu}$ is collinear. Therefore we
can safely disregard $k^2$ in the denominator, and the vertex is
unaffected.  

The collinear contribution comes from Fig.~\ref{fig2} (a), and the
amplitude is given by
\begin{equation}
M_{CO}  = -ig^2 C_F \int \frac{d^D l}{(2\pi)^D} 
\frac{2\overline{n}\cdot (l+p) \Gamma}{l^2 (l+p)^2 \overline{n}\cdot
  l}=\frac{\alpha_s C_F}{4\pi} \Gamma I_a (p^2),
\label{coll}
\end{equation}
where only the leading terms are collected by considering that the
loop momentum is collinear. And $I_a (p^2)$ is given by
\begin{eqnarray}
I_a (p^2) &=& \Bigl(\frac{-p^2}{\mu^2} \Bigr)^{-\epsilon} \Bigl(
\frac{2}{\epsilon^2} 
+\frac{2}{\epsilon} +4-\frac{\pi^2}{6} \Bigr) \nonumber \\
&=& \frac{2}{\epsilon^2} +\frac{2}{\epsilon} \Bigl( 1-\ln
\frac{-p^2}{\mu^2} \Bigr) -2\ln \frac{-p^2}{\mu^2} +\ln^2
\frac{-p^2}{\mu^2} +4-\frac{\pi^2}{6}.
\label{ia}
\end{eqnarray}
Here we neglected the terms which approach zero as $p^2 \rightarrow
0$. 

As a concrete example, let us show how the calculation of $M_{CO}$ can
be performed using the modified minimal subtraction
($\overline{\mathrm{MS}}$) scheme with the spacetime dimension
$D=4-2\epsilon$.  We first combine the denominator
using the Feynman parametrization technique as
\begin{equation}
\frac{1}{l^2 (l+p)^2
  \overline{n}\cdot l} = 4\int_0^1 dx \int_0^{\infty} du
  \frac{1}{(l+x p +u\overline{n} )^2 -\Bigl(2ux \overline{n}\cdot p -x
  (1-x)p^2\Bigr)}.
\end{equation}
Shifting the loop momentum as $l^{\mu} \rightarrow l^{\mu} - x
p-u\overline{n}^{\mu}$, $M_{CO}$ becomes
\begin{eqnarray}
M_{CO} &=& -2ig^2 C_F 4 \Gamma \Bigl(\frac{\mu^2 e^{\gamma}}{4\pi}
  \Bigr)^{\epsilon} \int_0^1 dx 
  \int_0^{\infty} du \int \frac{d^D 
  l}{(2\pi)^D} \frac{(1-x) \overline{n}\cdot p}{\Bigl(l^2
  -(2ux\overline{n}\cdot p - x(1-x)p^2)\Bigr)^3} \nonumber \\
&=& -\frac{\alpha_s C_F}{4\pi}  4\Gamma
 \Bigl(\frac{\mu^2 e^{\gamma}}{4\pi}
  \Bigr)^{\epsilon} \overline{n}\cdot p \Gamma 
  (1+\epsilon) \int_0^1 dx (1-x) \int_0^{\infty} du
  \Bigl(2ux\overline{n}\cdot p - x(1-x)p^2\Bigr)^{-1-\epsilon}
  \nonumber \\ 
&=& -\frac{\alpha_s C_F}{4\pi} 2\Gamma \Bigl(\frac{\mu^2
  e^{\gamma}}{4\pi}   \Bigr)^{\epsilon}
  (-p^2)^{-\epsilon} \frac{\Gamma(1+\epsilon)}{\epsilon} \int_0^1 dx
  (1-x)^{1-\epsilon} x^{-1-\epsilon} \nonumber \\
&=& \frac{\alpha_s C_F}{4\pi}\Gamma \Bigl(\frac{-p^2}{\mu^2}
  \Bigr)^{-\epsilon} \Bigl( \frac{2}{\epsilon^2} +\frac{2}{\epsilon}
  +4-\frac{\pi^2}{6} \Bigr) =  \frac{\alpha_s C_F}{4\pi} \Gamma
  I_a (p^2) \nonumber \\
&=& \frac{\alpha_s C_F}{4\pi} \Gamma\Bigl[ \frac{2}{\epsilon^2} +
  \frac{2}{\epsilon} \Bigl( 1-\ln \frac{-p^2}{\mu^2} \Bigr) -2\ln
  \frac{-p^2}{\mu^2} +\ln^2 \frac{-p^2}{\mu^2}
  +4-\frac{\pi^2}{6}\Bigr]. 
\label{collin}
\end{eqnarray}
The $1/\epsilon$ poles are ultraviolet divergences and the infrared
divergences appear as logarithms of $p^2$. If we put $p^2=0$, there is
no definite scale and the diagram becomes zero. The ultraviolet and
the infrared divergences cancel in this case, as is clear from
Eq.~(\ref{ia}). 

The usoft contribution is given by Fig.~\ref{fig2} (b), and it is
written as
\begin{eqnarray}
M_{US} &=& -ig^2 C_F \int \frac{d^D l}{(2\pi)^D} \frac{\overline{n}\cdot p 
\gamma^{\mu} (\FMslash{l}+\FMslash{k})\FMslash{n}}{l^2 (\overline{n}\cdot p 
n\cdot l +p^2) (l+k)^2} \nonumber \\
&=&\frac{\alpha_s C_F}{4\pi} \gamma^{\mu}
\Bigl[\frac{2}{\epsilon} \Bigl( 1- \ln \frac{q^2}{p^2} \Bigr) -2\ln
  \frac{-k^2}{\mu^2} -2\ln \frac{p^2}{q^2} \ln \frac{-k^2}{\mu^2}
  -\ln^2 \frac{p^2}{q^2} +4-\frac{2\pi^2}{3} \Bigr]. 
\label{soft}
\end{eqnarray}
Here the $1/\epsilon$ poles are ultraviolet divergences and the
infrared divergences appear as logarithms of $p^2$.

The remaining contributions from Fig.~\ref{fig2} (c) -- (f) are
hard-collinear contributions. The contribution from Fig.~\ref{fig2}
(c) is given by
\begin{equation}
M_c = -ig^2 C_F \Gamma
\int \frac{d^D l}{(2\pi)^D} \frac{2\overline{n}\cdot
  (l+p)}{l^2 (l^2 + \overline{n} \cdot p n\cdot l) \overline{n}
  \cdot l} = 0,
\label{mc}
\end{equation}
where the off-shellness $p^2$ is suppressed by
$\Lambda$ and it is discarded according to the power counting method
described earlier. If we disregard the power counting, we can imagine
performing the calculation with an arbitrary infrared regulator
in the denominator. Then there appears the dependence on the infrared
cutoff. But when we add all the hard-collinear contributions
Fig.~\ref{fig2} (c) -- (f), we have verified that this dependence on
the fictitious infrared cutoff is canceled, 
which means that the treatment of the off-shellness is consistent.

In Fig.~\ref{fig2} (d), there are two types of contributions in which
the currents $J_{\xi q}^{(1)}$, $J_c^{(1)}$ contribute, and the
currents $J_{\xi q}^{(2)}$, $J_c^{(0)}$ contribute. The Feynman
diagram in Fig.~\ref{fig2} (d) for the first case gives
\begin{eqnarray}
M_d^{(1,1)} &=& -ig^2 C_F \int \frac{d^D l}{(2\pi)^D}
  \frac{\overline{n} \cdot 
  (l+p) \overline{n} \cdot l}{l^2 (l^2 +\overline{n} \cdot p n\cdot l
  )(l^2 +\overline{n} \cdot l n\cdot k)} \nonumber \\
&\times& \Bigl[ \frac{\gamma_{\perp\alpha}
  \FMslash{l}_{\perp}}{\overline{n} \cdot (l+p)} \Bigl(
  \frac{\FMslash{l}_{\perp} \gamma_{\perp\mu}}{\overline{n} \cdot
  (l+p)} +\frac{\gamma_{\perp\mu}\FMslash{l}_{\perp}}{\overline{n}
  \cdot l} \Bigr) \gamma_{\perp}^{\alpha} -2 \Bigl(
  \frac{\FMslash{l}_{\perp} \gamma_{\perp\mu}}{\overline{n} \cdot 
  (l+p)} +\frac{\gamma_{\perp\mu}\FMslash{l}_{\perp}}{\overline{n}
  \cdot l} \Bigr) \frac{\FMslash{l}_{\perp}}{\overline{n} \cdot l}
  \Bigr] \nonumber \\
&=& -ig^2 C_F \int \frac{d^D l}{(2\pi)^D} \frac{1}{l^2 (l^2
  +\overline{n} \cdot p n\cdot l 
  )(l^2 +\overline{n} \cdot l n\cdot k)} \nonumber \\
&\times& \Bigl[(4-D) \frac{ (\overline{n} \cdot l) l_{\perp}^2
  \gamma_{\perp\mu}}{\overline{n} \cdot (l+p)} +(2-D)
  \FMslash{l}_{\perp} \gamma_{\perp\mu} \FMslash{l}_{\perp}
  -\frac{2\overline{n} \cdot (l+p) l_{\perp}^2
  \gamma_{\perp\mu}}{\overline{n} \cdot l} \Bigr],
\label{md11}
\end{eqnarray}
where the Dirac algebra for the perpendicular components in $D$
dimensions is given by
\begin{eqnarray}
\gamma_{\perp\mu}\gamma^{\mu}_{\perp} &=& D-2, \ \ \gamma_{\perp\mu}
\FMslash{a}_{\perp}\gamma^{\mu}_{\perp} = (4-D)\FMslash{a}_{\perp},
\nonumber \\
\gamma_{\perp\mu} \FMslash{a}_{\perp} \FMslash{b}_{\perp}
\gamma^{\mu}_{\perp} &=& 4 a_{\perp} \cdot b_{\perp} +(D-6)
\FMslash{a}_{\perp} \FMslash{b}_{\perp},  \nonumber \\
\gamma_{\perp\mu} \FMslash{a}_{\perp} \FMslash{b}_{\perp}
\FMslash{c}_{\perp} \gamma^{\mu}_{\perp} &=& -2 \FMslash{c}_{\perp}
\FMslash{b}_{\perp}  \FMslash{a}_{\perp} +(6-D) \FMslash{a}_{\perp}
\FMslash{b}_{\perp} \FMslash{c}_{\perp},
\end{eqnarray}
for arbitrary four-vectors $a^{\mu}$, $b^{\mu}$ and $c^{\mu}$. 
The first two terms are finite and each term gives $-\alpha_s C_F
\gamma_{\perp\mu}/(4\pi)$, and only the last term involves
divergences. In the last term, $l_{\perp}^2$ can be written as
\begin{equation}
l_{\perp}^2 = l_{\perp}^2 +\overline{n} \cdot l n\cdot (l+k)
-\overline{n} \cdot l n\cdot (l+k)  
= l^2 + \overline{n}\cdot l n\cdot k -\overline{n}\cdot l n\cdot
(l+k).
\end{equation}
With this decomposition, the last term in Eq.~(\ref{md11}) can be
written as
\begin{eqnarray} 
&&2ig^2 C_F  \int \frac{d^D l}{(2\pi)^D} \frac{1}{l^2 (l^2
  +\overline{n} \cdot p n\cdot l 
  )(l^2 +\overline{n} \cdot l n\cdot k)} \frac{\overline{n} \cdot
  (l+p) l_{\perp}^2 
  \gamma_{\perp\mu}}{\overline{n} \cdot l} \nonumber \\
&&= 2ig^2 C_F \gamma_{\perp\mu} \Bigl[ \int \frac{d^D l}{(2\pi)^D}
  \frac{\overline{n}\cdot (l+p)}{l^2 (l^2 +\overline{n} \cdot p n\cdot
  l ) \overline{n}   \cdot l} \nonumber \\
&&- \int \frac{d^D l}{(2\pi)^D} \frac{\overline{n} \cdot (l+p) n\cdot
  (l+k)}{l^2 (l^2 +\overline{n} \cdot p n\cdot l 
  )(l^2 +\overline{n} \cdot l n\cdot k)} \Bigr] = \frac{\alpha_s
  C_F}{4\pi} \gamma_{\perp\mu} I_b (q^2),
\label{mdt11} 
\end{eqnarray}
where the first term is zero and  $I_b (q^2)$ is the second term which
is given by
\begin{eqnarray}
I_b (q^2) &=& \Bigl(\frac{-q^2}{\mu^2} \Bigr)^{-\epsilon} \Bigl(
-\frac{2}{\epsilon^2} -\frac{3}{\epsilon} -6+\frac{\pi^2}{6} \Bigr)
\nonumber \\
&=& -\frac{2}{\epsilon^2} +\frac{1}{\epsilon} \Bigl( -3+2\ln
\frac{-q^2}{\mu^2}\Bigr) +3\ln \frac{-q^2}{\mu^2}  -\ln^2
\frac{-q^2}{\mu^2} -6+\frac{\pi^2}{6}.
\end{eqnarray}
Here $q^2 = (p-k)^2 \approx -\overline{n}\cdot p n\cdot k \sim
Q\Lambda$, neglecting terms of order $\Lambda^2$.
As a result, $M_d^{(1,1)}$ is given by 
\begin{equation}
M_d^{(1,1)} = \frac{\alpha_s C_F}{4\pi} \gamma_{\perp\mu} \Bigl( I_b
(q^2) -2\Bigr).
\end{equation}

The second case of Fig.~\ref{fig2} (d) is given as
\begin{eqnarray}
M_d^{(2,0)} &=& -ig^2 C_F \frac{\FMslash{\overline{n}}}{2} n_{\mu}
\int \frac{d^D l}{(2\pi)^D} \frac{1}{l^2 (l^2 \overline{n} \cdot p
      n\cdot l ) (l^2 + \overline{n} \cdot l n\cdot k)}\nonumber
\\
&&\times \Bigl( -\frac{2\overline{n} \cdot (l+p)
        l^2}{\overline{n} \cdot l} +(D-2) l_{\perp}^2 \Bigr).
\label{md200}
\end{eqnarray}
The integrand of the first term in Eq.~(\ref{md200}) can be written as
\begin{eqnarray}
&&  -\frac{2\overline{n} \cdot (l+p)
        \Bigl[(l^2+\overline{n}\cdot l n\cdot k) -\overline{n} \cdot l
        n\cdot (l+k) +\overline{n}\cdot l n\cdot l\Bigr]}{l^2 (l^2 +
        \overline{n} \cdot p 
       n\cdot l ) (l^2 + \overline{n} \cdot l n\cdot k) \overline{n}
        \cdot l} \nonumber \\
&&= -\frac{2\overline{n} \cdot (l+p)}{ l^2 (l^2 +\overline{n} \cdot p
        n \cdot l  )  \overline{n} \cdot l}
        +\frac{2\overline{n}\cdot (l+p) n\cdot (l+k)}{l^2 (l^2
        +\overline{n} \cdot p n\cdot l )( l^2 +\overline{n} \cdot
        l n\cdot k) } \nonumber \\
&&-\frac{2\overline{n} \cdot (l+p) n\cdot l}{l^2 (l^2 +\overline{n}
        \cdot p n\cdot l ) (l^2 +\overline{n} \cdot l n\cdot k)}.
\end{eqnarray}
Therefore by combining all these contributions, we have
\begin{eqnarray}
M_d^{(2,0)} &=& -ig^2 C_F \frac{\FMslash{\overline{n}}}{2} n_{\mu}
 \int \frac{d^D l}{(2\pi)^D}\Bigl[
 \frac{-2\overline{n} \cdot (l+p)}{
    l^2 (l^2 +\overline{n} \cdot p n\cdot l )  \overline{n} \cdot
    l} +  \frac{2\overline{n}\cdot (l+p) n\cdot
    (l+k)}{l^2 (l^2 +\overline{n} \cdot p n\cdot l )( l^2
    +\overline{n}  \cdot l n\cdot k) } \nonumber \\
&+&  \frac{ (D-2) l_{\perp}^2   -2\overline{n} \cdot (l+p) n\cdot 
    l}{l^2 (l^2 +\overline{n} 
        \cdot p n\cdot l ) (l^2 +\overline{n} \cdot l n\cdot k)}
  \Bigr] = \frac{\alpha_s C_F}{4\pi}  \frac{\FMslash{\overline{n}}}{2}
n_{\mu} \Bigl(  I_b (q^2) + I_c (q^2) \Bigr),
\label{md20} 
\end{eqnarray}
where the first term here also vanishes. And $I_c (q^2)$ comes from
the last term in Eq.~(\ref{md20}), which is given as
\begin{equation}
I_c (q^2) = 2 + \Bigl(\frac{-q^2}{\mu^2} \Bigr)^{-\epsilon}
\Bigl(\frac{2}{\epsilon} +5\Bigr) =
\frac{2}{\epsilon} -2\ln \frac{-q^2}{\mu^2} +3.  
\end{equation}

The Feynman diagram in Fig.~\ref{fig2} (e) is an additional diagram with
the insertion of $\mathcal{L}_c^{(1)}$, which also contributes at
leading order. It is given as
\begin{eqnarray}
M_e &=& 2ig^2 C_F n_{\mu} \int \frac{d^D l}{(2\pi)^D}  \frac{l_{\perp}
  \cdot k_{\perp} \overline{n} \cdot (l+p) \overline{n} \cdot l}{l^2
  (l^2 +\overline{n}\cdot p n\cdot l )(l^2 +\overline{n} \cdot l
  n\cdot k)^2} \nonumber \\
&&\times\Bigl[ (4-D) \overline{n} \cdot l -2\overline{n} \cdot
  (l+p) \Bigr] \FMslash{l}_{\perp}.
\end{eqnarray}
In the calculation of $M_e$, the result is proportional to
$\FMslash{k}_{\perp}$, which can be written as
\begin{equation}
\FMslash{k}_{\perp} = \FMslash{k} -\frac{\FMslash{n}}{2}
\overline{n}\cdot k -\frac{\overline{\FMslash{n}}}{2} n\cdot k.
\end{equation}
Since this is sandwiched between $\overline{\xi}$ and $q_{us}$,
$\FMslash{k}$ vanishes due do the equation of motion, and the part
with $\FMslash{n}$ also vanishes due to $\xi$. And the final result
can be written as
\begin{equation}
M_e = \frac{\alpha_s C_F}{4\pi} \frac{\overline{\FMslash{n}}}{2}
n_{\mu} \Bigl( -\frac{2}{\epsilon} +2\ln \frac{-q^2}{\mu^2} -5 \Bigr)=
\frac{\alpha_s C_F}{4\pi} \frac{\overline{\FMslash{n}}}{2} 
n_{\mu} \Bigl(-2-I_c (q^2) \Bigr).
\end{equation}
The Feynman diagram in Fig.~\ref{fig2} (f) is introduced because it is
of the same order as the other hard-collinear contributions, but it is
zero. 

When we sum over all the hard-collinear contribution from
Fig.~\ref{fig2} (c)--(f), the hard-collinear
contribution is given as 
\begin{eqnarray}
M_{HC}&=&M_c + M_d^{(1,1)} + M_d^{(2,0)} + M_e  \nonumber \\
&=& \frac{\alpha_s C_F}{4\pi} \gamma^{\mu}
\Bigl[-\frac{2}{\epsilon^2}
+\frac{1}{\epsilon} \Bigl( -3+\ln \frac{-q^2}{\mu^2} \Bigr) +3\ln
\frac{-q^2}{\mu^2} -\ln^2 \frac{-q^2}{\mu^2}  -8+\frac{\pi^2}{6}
\Bigr]. 
\label{mhc}
\end{eqnarray}
Here we use the fact that $\gamma_{\perp\mu} +
\FMslash{\overline{n}}n_{\mu}/2 = \gamma_{\mu} - \FMslash{n}
\overline{n}_{\mu}/2$, and the component with $\FMslash{n}$ vanishes
when attached to the collinear quark field.  
The hard-collinear contribution $M_{HC}$ looks fortuitously the same
as the result in full QCD with the on-shell renormalization scheme if
we disregard the origin of the $1/\epsilon$ poles. However, in the
result of the full theory in Eq.~(\ref{mfd}), only the first
$1/\epsilon$ pole is the ultraviolet divergence and the remaining
divergences are infrared. On the other hand, the divergences of the
hard-collinear contribution in Eq.~(\ref{mhc}) are ultraviolet
divergences. 

The infrared divergences in Eq.~(\ref{mhc}) is handled by
the off-shellness, but there is no need for the infrared cutoff for
the hard-collinear contributions from the power counting analysis. The
first terms in Eqs.~(\ref{mdt11}), (\ref{md20}) are those proportional
to $M_c$, which is zero. Even though it becomes nonzero, say, by
putting a fictitious infrared cutoff, these all cancel when we sum all
the hard-collinear contributions. As a result, the hard-collinear
contribution contains only the ultraviolet divergences.

If we add the collinear and the soft contributions in
Eqs.~(\ref{collin}), (\ref{soft}), we obtain
\begin{equation}
M_{CO} +M_{US} = \frac{\alpha_s C_F}{4\pi} \gamma_{\mu}
\Bigl[\Bigl(\frac{2}{\epsilon^2} 
+\frac{4}{\epsilon} \Bigr) 
\Bigl(\frac{-q^2}{\mu^2} \Bigr)^{-\epsilon} -2 \ln \frac{p^2}{q^2} \ln
\frac{k^2}{q^2} -2\ln\frac{p^2}{q^2} -2\ln
\frac{k^2}{q^2}+8-\frac{5\pi^2}{6}\Bigr], 
\end{equation}
from which we can see explicitly that the infrared divergences [$\ln
(p^2/q^2)$, $\ln (k^2/q^2)$ terms] are exactly the same as those from
the full theory given by Eq.~(\ref{irf}). Another interesting feature
of the calculation appears when we add all the hard-collinear,
collinear and soft contributions, which is given by
\begin{equation}
M_{HC}+M_{CO}+M_{US} = \frac{\alpha_s C_F}{4\pi} \gamma_{\mu} \Bigl(
\frac{1}{\epsilon} -  \ln \frac{-q^2}{\mu^2}-2 \ln
\frac{p^2}{q^2} \ln \frac{k^2}{q^2} -2 \ln \frac{p^2}{q^2} -2 \ln
\frac{k^2}{q^2} -\frac{2\pi^2}{3} \Bigr),
\label{sc1ir}
\end{equation}
which is exactly the same as the result from the full theory in
Eq.~(\ref{irf}). It means that, though the calculation in
$\mathrm{SCET}_{\mathrm{I}}$ is complicated due to the classification
of the effective Lagrangian and the current operators by power
counting, the calculation in $\mathrm{SCET}_{\mathrm{I}}$ corresponds
to the separation of the full-theory results into different kinematic
regions. In fact, there is no hard contribution of order $Q$ in the
full theory because the radiative corrections involve the relativistic
invariants, which are $p\cdot k\sim \mathcal{O} (Q\Lambda)$, $p^2, k^2
\sim \Lambda^2$. Therefore there is no change in the behavior of the
soft-collinear currents including their renormalization effects
near the boundary $\mu =Q$. Of course, if we consider, for example,
spectator effects including the soft-collinear currents as in $B$
decays, the situation will be completely different. 

\begin{figure}[b]
\hspace{-5.0cm}\vspace{0.5cm}
\epsfig{file=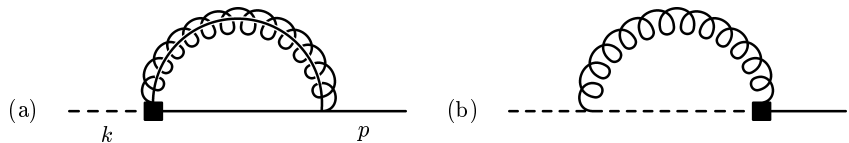, width=6.3cm}\vspace{-1.0cm}
\caption{Vertex correction to the soft-collinear current in
  $\mathrm{SCET}_{\mathrm{II}}$. (a) collinear contribution, (b) soft
  contribution.}
\label{fig6}
\end{figure}

The behavior of the soft-collinear current changes drastically in
$\mathrm{SCET}_{\mathrm{II}}$ because there are no hard-collinear
contributions, which are already integrated out in
$\mathrm{SCET}_{\mathrm{I}}$. There are only collinear and soft
contributions. In order to calculate these contributions, we evolve
the soft-collinear current down to $\mathrm{SCET}_{\mathrm{II}}$ and
compute the radiative corrections at one loop. There is no effect from
the effective Lagrangian in $\mathrm{SCET}_{\mathrm{II}}$ since the
inserted Lagrangian starts from $\alpha_s$, and it
does not contribute to the correction at order $\alpha_s$. Therefore
the necessary change is to replace $Y$ by $S$, and the Feynman graphs
contributing to the radiative correction of the soft-collinear current
in $\mathrm{SCET}_{\mathrm{II}}$ are shown in Fig.~\ref{fig6}.

The results of the collinear and soft contributions are the same as
those obtained in $\mathrm{SCET}_{\mathrm{I}}$, which are given by
\begin{eqnarray}
M_{CO} +M_S &=& \frac{\alpha_s C_F}{4\pi} \gamma_{\mu}
\Bigl[\Bigl(\frac{2}{\epsilon^2} 
+\frac{4}{\epsilon} \Bigr) 
\Bigl(\frac{-q^2}{\mu^2} \Bigr)^{-\epsilon} -2 \ln \frac{p^2}{q^2} \ln
\frac{k^2}{q^2} -2\ln\frac{p^2}{q^2} -2\ln
\frac{k^2}{q^2}+8-\frac{5\pi^2}{6}\Bigr] \nonumber \\
&=&\frac{\alpha_s C_F}{4\pi} \gamma^{\mu}
\Bigl[\frac{2}{\epsilon^2} +\frac{2}{\epsilon} \Bigl(2-\ln
  \frac{-q^2}{\mu^2} \Bigr) -4\ln   \frac{-q^2}{\mu^2}+\ln^2
  \frac{-q^2}{\mu^2} \nonumber \\
&&-2 \ln \frac{p^2}{q^2} \ln
\frac{k^2}{q^2} -2\ln\frac{p^2}{q^2} -2\ln
\frac{k^2}{q^2}+8-\frac{5\pi^2}{6}\Bigr].
\label{scet2}
\end{eqnarray}
Here again, the infrared divergence in $\mathrm{SCET}_{\mathrm{II}}$
is the same as that in the full theory, and in
$\mathrm{SCET}_{\mathrm{I}}$, which makes the matching between the
full theory and the effective theories possible. However the
ultraviolet behavior of the soft-collinear current is different from
the current in the full theory or $\mathrm{SCET}_{\mathrm{I}}$.  

If we calculate the collinear and the soft contributions to the
soft-collinear current in $\mathrm{SCET}_{\mathrm{II}}$ with external
particles on the mass shell ($p^2=k^2=0$) and use the dimensional
regularization for both the ultraviolet and the infrared divergences,
all the Feynman diagrams in 
Fig.~\ref{fig6} vanish and the infrared divergences are canceled by
the ultraviolet divergences since there are no scales
involved. However, since we know the ultraviolet divergences
explicitly in Eq.~(\ref{scet2}), we can read off the infrared and the
ultraviolet divergences in $\mathrm{SCET}_{\mathrm{II}}$ as
\begin{equation}
M_{CO} + M_S =\frac{\alpha_s C_F}{4\pi} \gamma^{\mu} \Bigl[
  \frac{2}{\epsilon_{\mathrm{UV}}^2} 
+\frac{2}{\epsilon_{\mathrm{UV}}} \Bigl( 2-\ln \frac{-q^2}{\mu^2}
\Bigr) -\frac{2}{\epsilon_{\mathrm{IR}}^2}
-\frac{2}{\epsilon_{\mathrm{IR}}} \Bigl( 2-\ln \frac{-q^2}{\mu^2}
\Bigr)\Bigr].
\label{scet2s}
\end{equation}

With all this information, we can calculate the Wilson coefficients
and the anomalous dimension of the (u)soft-collinear current at leading
order in SCET, and to leading-log order in $\alpha_s$. Since the
soft-collinear current is a conserved current in the full theory, it
is not renormalized. It can be explicitly seen by noting that the
$1/\epsilon_{\mathrm{UV}}$ pole in the vertex correction is canceled 
by the wave function renormalization. The radiative corrections in the
full theory and in $\mathrm{SCET}_{\mathrm{I}}$ are the same 
when we put the external particles on the mass shell [See
  Eqs.~(\ref{mfd}), (\ref{mhc}) and (\ref{scet2s}).], or off the mass
shell [See Eqs.~(\ref{irf}), (\ref{sc1ir}).], therefore the Wilson
coefficient is 1 to order $\alpha_s$ and the soft-collinear current
operator does not scale in the region $\sqrt{Q\Lambda} <\mu <Q$
either. However, in $\mathrm{SCET}_{\mathrm{II}}$, the ultraviolet
behavior becomes different and the Wilson coefficient for the
soft-collinear operator is given by
\begin{equation}
C_{\mathrm{II}} (\mu) = 1+\frac{\alpha_s (\mu) C_F}{4\pi} \Bigl( 3\ln
\frac{-q^2}{\mu^2} -\ln^2 \frac{-q^2}{\mu^2} -8+\frac{\pi^2}{6}\Bigr), 
\end{equation}
including the contribution from the wavefunction renormalization. Note
that the contribution of the wavefunction renormalization in the full
theory is the same for the collinear quark and the (u)soft
quark. Therefore they cancel in matching, but it is needed in
calculating the Wilson coefficient. The Wilson coefficient will be
matched at $\mu =\sqrt{-q^2} \sim \sqrt{Q\Lambda}$
\begin{equation}
C_{\mathrm{II}}(\sqrt{-q^2}) = 1+\frac{\alpha_s (\sqrt{-q^2})
  C_F}{4\pi}  \Bigl( -8+\frac{\pi^2}{6}\Bigr), 
\end{equation}
and it satisfies the renormalization group equation
\begin{equation}
\mu \frac{d}{d\mu} C_{\mathrm{II}} (\mu) =\gamma_{\mathrm{II}} (\mu)
C_{\mathrm{II}} (\mu).
\end{equation}
The counterterm $Z$ for the current operator is given by
\begin{equation}
Z=1+\frac{\alpha_s C_F}{4\pi} \Bigl[ \frac{2}{\epsilon^2}
  +\frac{1}{\epsilon} \Bigl( 3-2\ln \frac{-q^2}{\mu^2} \Bigr)\Bigr],
\end{equation}
including the counterterm from the wave function renormalization. The
anomalous dimension is given by
\begin{equation}
\gamma_{\mathrm{II}} = \frac{1}{Z} \Bigl( \mu
\frac{\partial}{\partial \mu} +\beta \frac{\partial}{\partial g}
\Bigr) Z 
= -\frac{\alpha_s C_F}{4\pi} \Bigl( 6-4\ln \frac{-q^2}{\mu^2}\Bigr).
\end{equation}

\section{\label{sec4} Comparison with other current operators}
With our understanding of the behavior of the (u)soft-collinear
current operators in SCET, we can analyze the behavior of other
current operators. Here we discuss the cases of the  back-to-back
collinear current and the heavy-to-light current. The back-to-back
collinear current has been considered in the four-quark operators for
nonleptonic $B$ decays \cite{Chay:2003zp,Chay:2003ju}, and in  
deep inelastic scattering in the endpoint region where SCET can be
applied \cite{Manohar:2003vb}. 
The scaling of the momentum components in deep inelastic
scattering is a little different from the power counting method
described here. We describe the behavior of the back-to-back collinear
current using two-step matching with the large scale $Q$, which
separates the full theory and $\mathrm{SCET}_{\mathrm{I}}$,
and the intermediate scale $\sqrt{Q\Lambda}$, which separates 
$\mathrm{SCET}_{\mathrm{I}}$ and $\mathrm{SCET}_{\mathrm{II}}$. Then
we go back to discuss the back-to-back current in deep inelastic
scattering.

It is useful to consider the characteristics of effective theories in
order to answer the question about whether a single-step matching and
a two-step matching will produce equivalent results. In constructing
an effective theory, we integrate out the degrees of freedom above
some large energy scale $Q$. Therefore the ultraviolet behavior is altered
in the effective theory compared to the full theory, while the
infrared behavior remains the same because two theories share the same
infrared region. Therefore, if there are some hard contributions of
order $Q$ for some operators in the full theory, they are not
reproduced in the effective theory. As a result, nontrivial
Wilson coefficients and anomalous dimensions are developed for that
operators in the effective theory in matching. 

In the context of SCET, if there are hard contributions of order $Q$
for an operator in the full theory, nontrivial Wilson coefficients and
anomalous dimensions for the operator appear in
$\mathrm{SCET}_{\mathrm{I}}$. It happens for the back-to-back
collinear current, and the heavy-to-light current. For the
back-to-back collinear current, the momenta of the external particles
are given by $p_n^{\mu} \sim Qn^{\mu}$ and $p_{\overline{n}}^{\mu}
\sim Q\overline{n}^{\mu}$. The radiative corrections contain
the relativistic invariant $p_n\cdot p_{\overline{n}} \sim Q^2$, which
are hard contributions. For the heavy-to-light current, the full
theory corresponds to the heavy quark effective theory for the heavy
quark with momentum $p_h^{\mu} \sim Q v^{\mu}$ and the full QCD for
the light quark with momentum $p_c^{\mu} \sim Qn^{\mu}$. The
radiative corrections contain the relativistic invariant of order 
$p_h \cdot p_c \sim Q^2$, which are hard contributions. Therefore
there is a nontrivial matching for the back-to-back collinear current
and the heavy-to-light current between the full theory and
$\mathrm{SCET}_{\mathrm{I}}$. 

It is possible to
have no hard contributions of order $Q$ in the full theory for some
reason, but if there are hard-collinear contributions of order
$\sqrt{Q\Lambda}$, the ultraviolet behavior of the full theory is the
same as that of $\mathrm{SCET}_{\mathrm{I}}$, which is given by the
hard-collinear contributions in $\mathrm{SCET}_{\mathrm{I}}$. This
happens for the usoft-collinear current. In this case the collinear
particle has momentum $p_c \sim Qn^{\mu}$ and the usoft particle has
$p_{us}^{\mu} \sim (\Lambda,\Lambda,\Lambda)$, where the largest
relativistic invariant is $p_c\cdot p_{us} \sim Q\Lambda$. The
radiative corrections in the full theory are hard-collinear
contributions and there are no hard contributions of order
$Q$. There should be nontrivial hard-collinear
contributions for the usoft-collinear current in
$\mathrm{SCET}_{\mathrm{I}}$, and they should give the same
contributions as in the full theory. Therefore the ultraviolet
behavior is the same in the full theory and in
$\mathrm{SCET}_{\mathrm{I}}$, and the matching between these theories
become trivial. The Wilson coefficients and the anomalous
dimensions are the same as those in the full theory. In other words,
there is no change of the behavior of the operators across the
boundary between the full theory and $\mathrm{SCET}_{\mathrm{I}}$. 

The criterion for nontrivial matching also applies to the matching
between $\mathrm{SCET}_{\mathrm{I}}$ and $\mathrm{SCET}_{\mathrm{II}}$.
If there are hard-collinear contributions of order
$\sqrt{Q\Lambda}$ in $\mathrm{SCET}_{\mathrm{I}}$, the ultraviolet
behavior of some operators in $\mathrm{SCET}_{\mathrm{I}}$ and
$\mathrm{SCET}_{\mathrm{II}}$ is different because the hard-collinear
modes are not present in $\mathrm{SCET}_{\mathrm{II}}$. And there is a
nontrivial matching between
$\mathrm{SCET}_{\mathrm{I}}$ and $\mathrm{SCET}_{\mathrm{II}}$. The
soft-collinear current belongs to this class, as explicitly shown in
this paper.

On the other hand, if there are no hard-collinear
contributions of order $\sqrt{Q\Lambda}$, the behavior of the
operators across the boundary between
$\mathrm{SCET}_{\mathrm{I}}$ and $\mathrm{SCET}_{\mathrm{II}}$ is
the same. This is true for the back-to-back collinear current and the
heavy-to-light currents. For the back-to-back collinear current, there
is no hard-collinear interaction between the quarks 
moving in the opposite directions, and the hard-collinear interaction
of the quarks with the hard-collinear gluons is given by the same form
as Eq.~(\ref{mc}), which is zero. 
\begin{figure}[t]
\hspace{-5.0cm}\epsfig{file=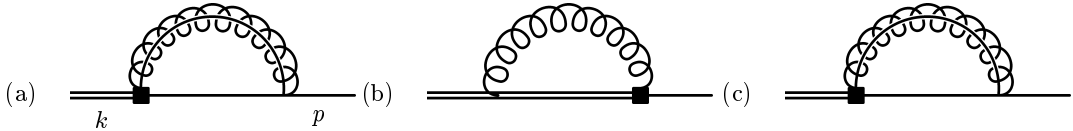, width=10.0cm}\vspace{-1.0cm}
\caption{Radiative corrections to the heavy-to-light current in
  $\mathrm{SCET}_{\mathrm{I}}$.  The
  double line is a heavy quark. (a) hard-collinear contribution, (b)
  usoft contribution, (c) collinear contribution.}
\label{fig7}
\end{figure}
For the heavy-to-light current, the radiative corrections to the
heavy-to-light current in $\mathrm{SCET}_{\mathrm{I}}$ is shown in
Fig.~\ref{fig7}. Here the hard-collinear contribution in
Fig.~\ref{fig7} (a) is zero. However, if there are
additional hard-collinear contributions, such as the spectator quark
interactions for $B$ decays, a nontrivial matching occurs as we go
down to $\mathrm{SCET}_{\mathrm{I}}$ and
$\mathrm{SCET}_{\mathrm{II}}$.

\begin{table}[b]
\begin{tabular}{c|ccc} \hline
Currents & (u)soft-collinear & back-to-back&
heavy-to-light\\ \hline
hard ($Q$)& X&O&O \\
hard-collinear ($\sqrt{Q\Lambda}$)&O&X&X \\ \hline
\end{tabular}
\caption{\label{table2}  Hard ($\sim Q$), and
  hard-collinear ($\sim \sqrt{Q\Lambda}$) contributions for various
  current operators.}  
\end{table}

We summarize the types of radiative corrections for the three current
operators in Table~\ref{table2}. From
Table~\ref{table2}, nontrivial matching between the full theory and
$\mathrm{SCET}_{\mathrm{I}}$ occurs for the back-to-back collinear
current and the heavy-to-light current,
and the (u)soft-collinear current needs nontrivial matching between
$\mathrm{SCET}_{\mathrm{I}}$ and
$\mathrm{SCET}_{\mathrm{II}}$. Therefore even though we apply the
two-step matching for these operators, one of the matching processes
is irrelevant because there is no change across the boundary, and only
one nontrivial matching appears. However, if there is an operator
which has hard and hard-collinear contributions in the full theory,
and hard-collinear contributions in $\mathrm{SCET}_{\mathrm{I}}$, the
two-step matching is necessary and there is a nontrivial matching
procedure at each boundary. An example of the operator which needs the
two-step matching appears in the four-quark operators for nonleptonic
$B$ decays. If a four-quark operator consists of a heavy quark, a soft
quark, and two collinear quarks in the opposite directions, the
interactions between the two collinear quarks and between a collinear
quark and the heavy quark produce hard contributions of order $Q$. The
interaction between a collinear quark and a soft quark produces
hard-collinear contributions of order $\sqrt{Q\Lambda}$. Therefore we
need two-step matching in this case.

There will be neither hard, nor hard-collinear
contributions in the radiative corrections of the (u)soft-(u)soft
current operator in the full theory and in
$\mathrm{SCET}_{\mathrm{I}}$ though we have not verified
explicitly. This is based on the fact that all the 
relativistic invariants in this case are of order $\Lambda^2$ in all
the theories. Then the matching will be 
trivial without any change of the behavior of the operator from the
full theory down to $\mathrm{SCET}_{\mathrm{II}}$, but it should be
checked explicitly.

The situation in deep inelastic scattering is subtler than the
discussion presented above. By choosing different reference 
frames, different types of hadronic current operators are
considered \cite{Manohar:2003vb}. In the target rest frame, the
relevant current is the (u)soft-collinear current operator, and in the
Breit frame where the photon has no energy component, the
corresponding current is the back-to-back collinear current
operator. 

In the target rest 
frame, the incoming quark has momentum $p^{\mu} = (\overline{n}\cdot
p, p_{\perp}, n\cdot p) \sim (\Lambda,\Lambda,\Lambda)$, which
is (u)soft. And the final-state quark has momentum $p_X^{\mu} \sim
(Q^2/\Lambda, \Lambda, (1-x) \Lambda)$ with $p_X^2 \sim Q^2 (1-x)$. In
this case, the boundary which separates the full theory and
$\mathrm{SCET}_{\mathrm{I}}$ is $\overline{n}\cdot p_X \sim
Q^2/\Lambda$, and the boundary 
between $\mathrm{SCET}_{\mathrm{I}}$ and $\mathrm{SCET}_{\mathrm{II}}$
is $\mu^2 \sim \overline{n}\cdot p_X n\cdot p \sim Q^2$. Therefore the
matching near $\mu=Q$ corresponds to the matching  
between $\mathrm{SCET}_{\mathrm{I}}$ and
$\mathrm{SCET}_{\mathrm{II}}$. As explained before, from very high
energy to $Q^2/\Lambda$ (from full theory to
$\mathrm{SCET}_{\mathrm{I}}$), there is no scaling of the
soft-collinear current.  
In the Breit frame, $p^{\mu} \sim (\Lambda^2/Q,\Lambda,Q)$, and
$p_X^{\mu} \sim (Q,\Lambda,Q(1-x))$ with $p_X^2 \sim Q^2 (1-x)$. Here
the boundary between the full theory and $\mathrm{SCET}_{\mathrm{I}}$
is $\overline{n}\cdot p_X\sim Q$ and the boundary between
$\mathrm{SCET}_{\mathrm{I}}$ and $\mathrm{SCET}_{\mathrm{II}}$ is
$\mu^2 \sim Q^2 (1-x)$. Therefore the matching near $\mu=Q$
corresponds to the matching between the full theory and
$\mathrm{SCET}_{\mathrm{I}}$. Therefore the matching procedures in
different reference frames correspond to the matching between
different effective theories. But the starting scale where nontrivial
evolution occurs is $\mu=Q$ in both matching. Therefore the Wilson
coefficients and the anomalous dimension are the same in both frames
just below $Q$. This interesting result
arises from the combination of the behavior of the soft-collinear
current and the back-to-back collinear current in
$\mathrm{SCET}_{\mathrm{I}}$ and $\mathrm{SCET}_{\mathrm{II}}$, and
the different scales which separate these effective theories.

\section{\label{sec5}Conclusion}
We have constructed the effective Lagrangian for
$\mathrm{SCET}_{\mathrm{I}}$ and $\mathrm{SCET}_{\mathrm{II}}$, which
are gauge invariant under collinear and (u)soft gauge
transformations at each order. It has been achieved in
$\mathrm{SCET}_{\mathrm{I}}$ by treating the usoft field as a
background field for collinear gauge transformations, and by
redefining the collinear gauge field. Also the usoft interactions with
the collinear particles are factorized. The effective Lagrangian in
$\mathrm{SCET}_{\mathrm{II}}$ is obtained by integrating out the
hard-collinear degrees of freedom in $\mathrm{SCET}_{\mathrm{I}}$.
This is the starting point to investigate various effects of the
interactions in $\mathrm{SCET}_{\mathrm{II}}$, which is the
final effective theory for systems with $p^2 \sim \Lambda^2$. For
example, at higher orders in $\Lambda$ and in $\alpha_s$, there are
interactions which violate  factorization properties in $B$ decays.
For many processes, these factorization-violating interactions are
subleading, but there may be interesting cases in which these
interactions can enter as the leading contribution. This will be
investigated in detail in a future publication.  

There are many points yet to be clarified in the effective Lagrangian
in $\mathrm{SCET}_{\mathrm{II}}$. We have constructed the effective
Lagrangian explicitly gauge invariant, but the momentum conservation
is not explicitly realized. Therefore, in the current form, we have to
consider if a certain vertex conserves collinear and soft momenta. If
not, it is not allowed. It would be desirable to have the effective
Lagrangian which is explicitly gauge invariant and momentum conserving
at each order in $\Lambda$ and $g$. If we choose a specific gauge
($\hat{W}=S=1$), the momentum conservation can be transparent, but we
also have to consider the corresponding change of the gauge fixing
Lagrangian. This will be discussed in more detail elsewhere. 

We have considered the radiative corrections of the (u)soft-collinear
current operators both in $\mathrm{SCET}_{\mathrm{I}}$ and in
$\mathrm{SCET}_{\mathrm{II}}$ at one loop and found that we can
reproduce the infrared divergence of the full theory in
$\mathrm{SCET}_{\mathrm{I}}$ and 
$\mathrm{SCET}_{\mathrm{II}}$ only in terms of the collinear, usoft
modes, and the collinear, soft modes respectively and we do not need
additional degrees of freedom. In regulating the infrared divergence
with the off-shellness, some modifications are needed for the Feynman
rules. But these are relatively easy and we have not invented more
complicated regularization schemes \cite{Beneke:2003pa,Bauer:2003td}
except the power counting method in expressing the propagators for
various cases.  

As we have seen in the behavior of the soft-collinear, back-to-back
collinear and heavy-to-light current operators, the presence of the
hard contributions of order $Q$ in the full theory, and the 
hard-collinear contributions of order $\sqrt{Q\Lambda}$ in
$\mathrm{SCET}_{\mathrm{I}}$ is important in determining which
matching gives a nontrivial result. It gives a criterion for using a
single-step matching or a two-step matching for the radiative
corrections. However, this is determined only at the end of the
calculation, and there is no a priori criterion.

\section*{Acknowledgments}
This work was supported by Korea Research Foundation Grant
(KRF-2003-041-C00052).

\end{document}